\documentclass[10pt,twocolumn,twoside]{IEEEtran}  

\IEEEoverridecommandlockouts                              

\usepackage{url}

\usepackage{amsmath,amssymb,amsthm}
\usepackage{mathtools}
\usepackage{tabularx,array,threeparttable}
\usepackage{multirow}
\usepackage{float}
\usepackage[pdftex]{graphicx}
\usepackage{wrapfig}
\usepackage{graphicx}
\usepackage{tikz}
\usetikzlibrary{automata,arrows,positioning,calc} 
\usepackage{multicol}
\usepackage{enumitem}
\setlist{leftmargin=5.5mm}
\usepackage{pgfplots}
\usepgfplotslibrary{fillbetween}	
\usepackage{xcolor}
\usepackage{colortbl}
\usepackage[noadjust]{cite}
\usepackage{booktabs}
\usepackage{hyperref}

\newcommand{\subparagraph}{}  
\usepackage{titlesec}
\titlespacing*{\subsection}{0pt}{0.15\baselineskip}{0.05\baselineskip}
\titlespacing*{\section}{0pt}{0.6\baselineskip}{0.5\baselineskip}

\newtheorem{theorem}{Theorem}
\newtheorem{lemma}{Lemma}
\newtheorem{definition}{Definition}
\newtheorem{remark}{Remark}

\newtheorem{assumption}{Assumption}

\newcommand{\argmin}{\operatornamewithlimits{arg\,min}}
\newcommand{\argmax}{\operatornamewithlimits{arg\,max}}

\newcommand{\demandhum}{\lambda^\text{h}}
\newcommand{\demandaut}{\lambda^\text{a}}

\newcommand{\pathidx}{i}
\newcommand{\numpaths}{N}
\newcommand{\pathset}{{[}\numpaths{]}}

\newcommand{\useridx}{j}

\newcommand{\congprof}{s}	
\newcommand{\congprofvec}{\boldsymbol{s}}	

\newcommand{\ffvelocity}{\bar{v}}
\newcommand{\numlanes}{b}
\newcommand{\spacehuman}{h^\text{h}}
\newcommand{\spaceaut}{h^\text{a}}

\newcommand{\denshum}{n^\text{h}}
\newcommand{\densaut}{n^\text{a}}
\newcommand{\flow}{f}
\newcommand{\flowhum}{f^\text{h}}
\newcommand{\flowaut}{f^\text{a}}
\newcommand{\flowhumvec}{\boldsymbol{f}^\text{h}}
\newcommand{\flowautvec}{\boldsymbol{f}^\text{a}}
\newcommand{\flowhumvecEQ}{\boldsymbol{f}^\text{h,EQ}}
\newcommand{\flowautvecEQ}{\boldsymbol{f}^\text{a,EQ}}

\newcommand{\jamden}{\bar{n}}
\newcommand{\critdens}{\tilde{n}}
\newcommand{\capacity}{\bar{F}}

\newcommand{\autlev}{\alpha}

\newcommand{\latency}{\ell}
\newcommand{\latencyvec}{\boldell}

\newcommand{\fflatency}{a}

\newcommand{\price}{p}
\newcommand{\pricevec}{\boldsymbol{p}}
\newcommand{\dominatedroads}{D}
\newcommand{\nondominatedroads}{\bar{\dominatedroads}}
\newcommand{\reward}{r}

\newcommand{\dataset}{\mathcal{D}}
\newcommand{\flowautnocontrol}{f^\text{b}}
\newcommand{\flowautnocontrolvec}{\boldsymbol{f}^\text{b}}

\newcommand{\midcs}{{[}m{]}}
\newcommand{\mprimeidcs}{{[}m'{]}}

\newcommand{\bs}{\boldsymbol{s}}
\newcommand{\bp}{\boldsymbol{p}}
\newcommand{\boldell}{\boldsymbol{\ell}}

\newcommand{\meq}{{m_\mathrm{EQ}}}
\newcommand{\meqidcs}{{{[}m_\mathrm{EQ}{]}}}
\newcommand{\mstareq}{{m^{*}_\mathrm{EQ}}}
\newcommand{\mstareqidcs}{{{[}m^{*}_\text{EQ}}{]}}
\newcommand{\mall}{{m_{\mathrm{ALL}}}}
\newcommand{\mstarall}{{m^{*}_{\mathrm{ALL}}}}
\newcommand{\mallidcs}{{{[}m_{\mathrm{ALL}}{]}}}
\newcommand{\altruismArg}{\kappa}
\newcommand{\altruismUnif}{\kappa_0}
\newcommand{\altruismFn}{\varphi}
\newcommand{\altruismLevSet}{K}
\newcommand{\eqDelay}{\hat{\ell}_0}

\DeclarePairedDelimiter\norm{\lVert}{\rVert}%
\newcommand\compactdots{\hbox to 1em{.\hss.\hss.}}
\newcommand{\asymeq}{\stackrel{\boldsymbol{\cdot}}{=}}

\usepackage[font=footnotesize]{caption}

\usepackage{xcolor}
\newcommand{\ignore}[1]{}

\title{\LARGE \bf
	Incentivizing Efficient Equilibria in \\ Traffic Networks with Mixed Autonomy
}

\author{Erdem B\i y\i k$^*$, Daniel A.~Lazar$^*$, Ramtin~Pedarsani, Dorsa~Sadigh
    \thanks{$^{*}$Authors contributed equally.}
	\thanks{Erdem B\i y\i k and Dorsa Sadigh are with the Department of Electrical Engineering, Stanford University. Dorsa Sadigh is also with the Department of Computer Science, Stanford University. Daniel Lazar and Ramtin Pedarsani are with the Department of Electrical and Computer Engineering, UC Santa Barbara. {\tt\small ebiyik@stanford.edu}, {\tt\small dlazar@ece.ucsb.edu}, {\tt\small ramtin@ece.ucsb.edu}, {\tt\small dorsa@stanford.edu}}
	\thanks{This work was supported by NSF ECCS grant \#1952920, FLI grant RFP2-000, and Toyota.}
}

\begin{document}
	
\maketitle
\thispagestyle{empty}
\pagestyle{empty}

\begin{abstract}
	
Traffic congestion has large economic and social costs. The introduction of autonomous vehicles can potentially reduce this congestion by increasing road capacity via vehicle platooning and by creating an avenue for influencing people's choice of routes. We consider a network of parallel roads with two modes of transportation: (i) human drivers, who will choose the quickest route available to them, and (ii) a ride hailing service, which provides an array of autonomous vehicle route options, each with different prices, to users. We formalize a model of vehicle flow in mixed autonomy and a model of how autonomous service users make choices between routes with different prices and latencies. Developing an algorithm to learn the preferences of the users, we formulate a planning optimization that chooses prices to maximize a social objective. We demonstrate the benefit of the proposed scheme by comparing the results to theoretical benchmarks which we show can be efficiently calculated.
\end{abstract}

\section{INTRODUCTION}

Road congestion is a major and growing source of inefficiency, costing drivers in the United States billions of dollars in wasted time and fuel \cite{schrank2015}. Autonomous vehicles could improve the efficiency of road usage by forming platoons, thereby increasing road capacity and smoothing traffic flow. The presence of \emph{mixed autonomy}, where human drivers and autonomous vehicles share roads, complicates these improvements -- if drivers make their routing decisions \emph{selfishly} and seek to minimize their experienced latency, this can result in suboptimal network performance \cite{koutsoupias:1999fs}. Moreover, increasing road capacity by converting human-driven vehicles to autonomous ones can paradoxically worsen average transit user latency \cite{mehr2018can}.

It is therefore important to consider the effect of self-interested users on the system as well as how to \emph{influence} these users to take routes which mitigate this effect. In this paper we consider monetary incentivizes for such prosocial behavior. We consider a setting where a benevolent ride hailing service or social planner sets prices for each route that an autonomous vehicle user may take. The users of the autonomous service will choose their routes, or choose not to travel, based on their time-money tradeoff. The remaining human drivers will choose routes that minimize their latency. The role of the social planner is then to choose prices to optimize some social objective. Our model is an indirect Stackelberg game -- a game in which a social planner controls some fraction of the population's actions, where the remainder of the population responds selfishly. However, in our model the planner only controls its portion of the vehicle flow indirectly, via pricing.

To effectively set prices, the social planner needs a model for a) the flow of vehicles on a road, which depends on how many autonomous and human-driven vehicles are on the road, and b) how people make decisions between routes with various prices and latencies. We model the latter based on multinomial logits \cite{ben1985discrete} and model human drivers as selfish agents who reach a \emph{Nash Equilibrium} (also called a Wardrop Equilibrium in the context of transportation networks), a configuration in which no one can decrease their travel time by switching paths \cite{dafermos1980variationalinequalities}. 

Moreover, we use active preference-based learning \cite{sadigh2017active}, via a series of queries, to understand the preferences of autonomous service users. This enables the planner to predict how autonomous service users will react to a set of options with only a relatively low number of training queries. We experimentally verify this method accurately predicts human choice, and our planning algorithm can use this to improve network performance. Furthermore, we provide a theoretical framework which establishes benchmarks for the performance of the algorithm and we show that these benchmarks can be calculated in polynomial time. Finally, we show the efficacy of our pricing scheme in the context of these benchmarks.

Our contributions in this work are as follows:
\begin{itemize}[nosep]
\item We develop a formal mixed autonomy traffic flow model.
\item We develop an active preference-based algorithm to learn how different people value time and money in choosing their transportation option. This enables learning a model for people's routing choices in a data-efficient manner.
\item We use these models to formulate and solve an optimization for ride hailing service to minimize congestion and maximize the road network utilization while constraining a minimum profit for the service supplier.
\item We provide theoretical benchmarks for understanding the performance of the above and prove that we can calculate these benchmarks in polynomial time.  
\item We validate the learning algorithm and the planning optimization formulation via simulations and user studies. Our results show carefully designed pricing schemes can significantly improve traffic throughput.
\end{itemize}

\noindent\textbf{Related work.} Previous works have shown the potential benefits that autonomous vehicles can have for traffic networks by increasing road capacity through platooning \cite{lioris2017platoons, askari2017effect}, damping shockwaves of slowing vehicles \cite{stern2018dissipation, wu2018stabilizing}, managing merges \cite{jin2018modeling}, decongesting freeways in the event of accidents \cite{sivaranjani2015localization,lazar2019learning}, and balancing a supply of vehicles \cite{salazar2019congestion}. Relatedly, many works analyze and bound the inefficiency that can arise from network users choosing their routes selfishly  \cite{roughgarden2002bad}, including when autonomous vehicles are introduced \cite{mehr2018can}. 

As we use financial incentives to influence this behavior, our formulation is related to work in tolling, some of which consider users with different sensitivities to tolls \cite{fleischer2004tolls, brown2017robustness}. \cite{sandholm2002evolutionary} considers a congestion game framework and derives tolls which drive users to choose socially optimal strategies for a broad class of user strategy update dynamics. In contrast to many of these works, we consider an empirically validated probabilistic model for human choice \cite{daw2006cortical} which incorporates differences in price sensitivity, and we model vehicle flow on shared roads based on the foundational Fundamental Diagram of Traffic (FDT) \cite{daganzo1994cell}. Some works consider heterogeneous and stochastic user utilities (\emph{e.g.} \cite{koster2018preference}, \cite{sumalee2011first}). However, the approaches taken in these works cannot incorporate a FDT-based model for road congestion, in which latency is no longer an increasing function of vehicle flow on a road. Also relatedly, \cite{krichene2017stackelberg} considers a Stackelberg game (meaning a planner can route a portion of the vehicle flow) on parallel roads with flow dictated by the FDT when there is a single vehicle type. Hence, a major novelty of our work lies in that we consider a mixed-autonomy network where autonomous service users have different preferences.

To understand human choice, there has recently been much effort on learning human reward functions which are assumed to be sufficient to model preferences. Inverse reinforcement learning \cite{ng2000algorithms, abbeel2005exploration, ziebart2008maximum} and preference-based learning \cite{sadigh2017active, cheng2011preference,kallus2016revealed,zadimoghaddam2012efficiently} are the most popular choices. In this paper, we employ preference-based learning, a natural fit to our problem. We actively synthesize queries -- a non-trivial generalization and extension of \cite{sadigh2017active} -- for data-efficiency and better usability.

\section{PROBLEM SETTING AND OBJECTIVE}\label{sct:model}
\subsection{Vehicle Flow Model}
\label{subsec:vehicle_flow}

We assume every road $\pathidx$ has a maximum flow. This occurs when traffic is in \emph{free-flow} -- when all vehicles travel at the nominal road speed $\ffvelocity_\pathidx$.
\begin{definition}
The \emph{free-flow latency} of a road $\pathidx$, denoted $\fflatency_\pathidx$, is the time it takes vehicles to traverse the road in free-flow. With road length denoted $d_\pathidx$, the free-flow latency is $\fflatency_\pathidx := d_\pathidx/\ffvelocity_\pathidx$.
\end{definition}

\begin{figure}
	\centering
	\includegraphics[width=\columnwidth]{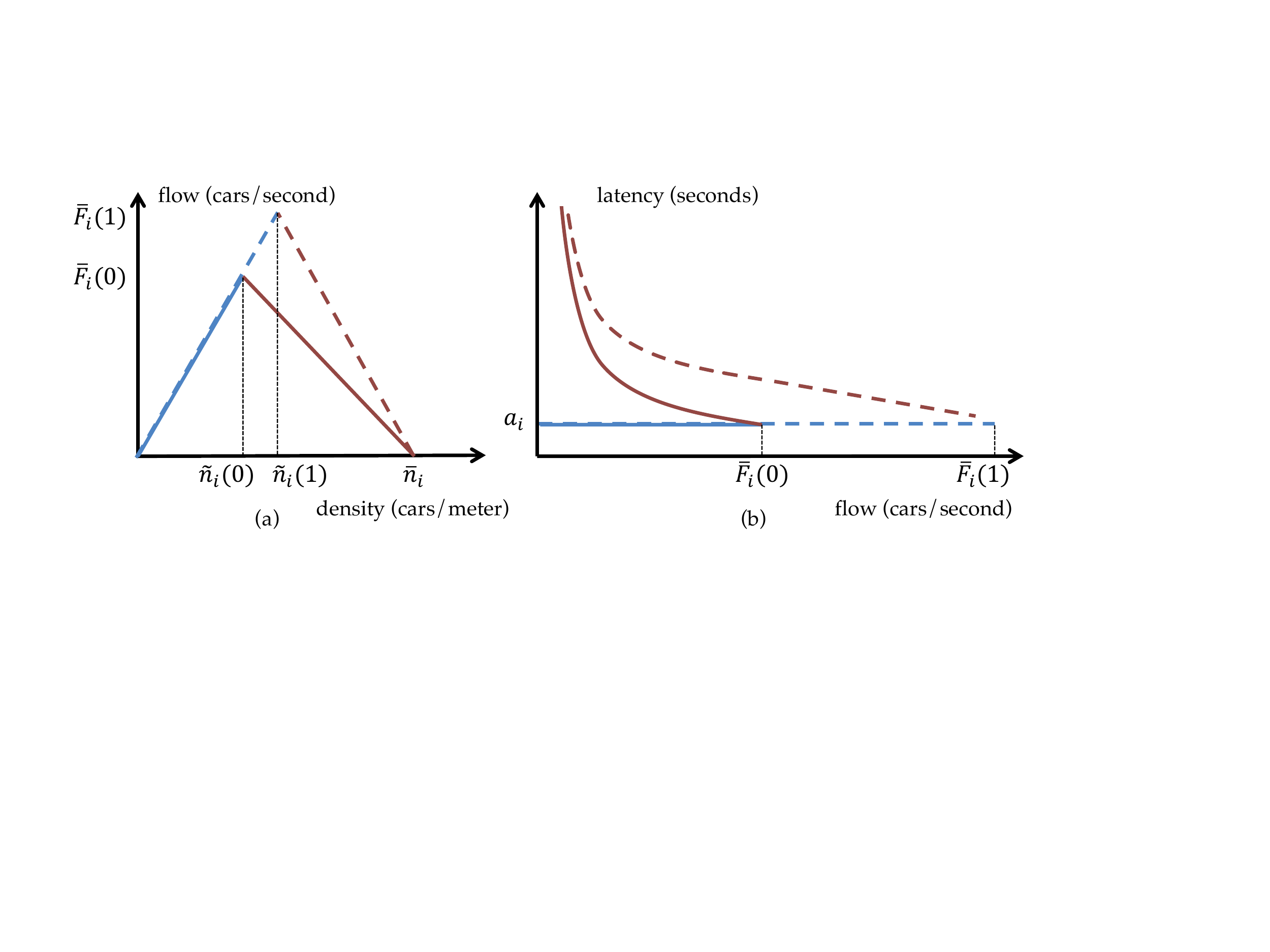}
	\vspace{-15px}
	\caption{\textbf{(a)} The Fundamental Diagram of Traffic for roads with all human-driven (solid) and all autonomous (dashed) vehicles. In the latter, congestion begins at a higher vehicle density as autonomous vehicles require a shorter headway when following other vehicles. \textbf{(b)} The relationship between vehicle flow and latency also changes in the presence of autonomous vehicles. Free-flow speed remains the same but maximum flow on a road increases.}
	\label{fig:mixed_fund_diag_and_laten}
	\vspace{-15px}
\end{figure}

\noindent
\textbf{Traffic Density vs Traffic Flow:} Adding more cars to a road that is already at maximum flow makes the traffic switch from free-flow to a congested regime, which decreases the vehicle flow. In the extreme case, at a certain density $\jamden_\pathidx$, cars are bumper-to-bumper and vehicle flow stops. 
The solid lines in Fig.~\ref{fig:mixed_fund_diag_and_laten}(a) -- Fundamental Diagram of Traffic~\cite{daganzo1994cell} -- illustrates this phenomenon, where flow increases linearly with respect to density until it hits the critical density. The slope corresponds to the free-flow velocity $\ffvelocity_\pathidx$ on road $i$. After the critical point, flow decreases linearly until it is zero at the maximum density. In this figure, $\capacity$ is the maximum flow and $\critdens$ denotes the critical density, where the argument is the fraction of vehicles that are autonomous. We will formally define this notation below.

\noindent
\textbf{Traffic Flow vs Road Latency:} The relationship between vehicle flow and road latency reflects the same free-flow/congested divide. As mentioned above, roads in free-flow have constant latency. In the congested regime, however, latency \emph{increases} as vehicle flow decreases, since a high density of vehicles is required to achieve a low traffic flow. This is represented in Fig.~\ref{fig:mixed_fund_diag_and_laten}(b) with the solid lines.


\noindent
\textbf{Mixed-Autonomy Roads.} We assume that on mixed-autonomy roads, the autonomous vehicles can coordinate with one another and potentially form platoons to help with the efficiency of the road network.
We now extend the traffic model above to mixed-autonomy settings as shown with dashed lines in Fig.~\ref{fig:mixed_fund_diag_and_laten}.
We define the autonomy level of a road $\pathidx$ as the fraction of autonomous vehicle flow on that road: $\autlev_\pathidx:=\frac{\flowaut_\pathidx}{\flowaut_\pathidx+\flowhum_\pathidx}$ where $\flowaut_\pathidx$ and $\flowhum_\pathidx$ represent the autonomous and human-driven vehicle flow, respectively. Assuming that neither the nominal velocity $\ffvelocity_\pathidx$ nor the maximum density $\jamden_\pathidx$ changes, the critical density at which traffic becomes congested will now shift and increase with \emph{autonomy level} $\autlev_\pathidx$, as platooned autonomous vehicles require a shorter headway than human drivers.

To formalize the relationship between autonomy level and critical density on road $i$, we assume the space occupied by autonomous vehicles and humans at nominal velocity is $\spaceaut_\pathidx$ and $\spacehuman_\pathidx$, respectively, with $\spaceaut_\pathidx\!\le\!\spacehuman_\pathidx$. This inequality reflects the assumption that autonomous vehicles can maintain a short headway, regardless of the type of vehicle they are following. Then, the critical density is
\begin{align}\label{eq:crit_dens}
\critdens_{\pathidx}(\autlev_{\pathidx}) := \frac{ \numlanes_{\pathidx}}{\autlev_{\pathidx}\spaceaut_\pathidx + (1-\autlev_{\pathidx})\spacehuman_\pathidx} \; ,
\end{align}
where $\numlanes_{\pathidx}$ is the number of lanes on that road. Here the denominator represents the average length from one car's rear bumper to the preceding car's rear bumper when all cars follow the vehicle in front of them with nominal headway. Note that critical density is expressed here as a function of the autonomy level $\autlev_\pathidx$ of the road. Since flow increases linearly with density until hitting the critical point, the maximum flow can also be expressed as a function of autonomy level: $\capacity_\pathidx(\autlev_\pathidx)=\ffvelocity_\pathidx \critdens_\pathidx(\autlev_{\pathidx}).$

The flow on a road, $\flow_\pathidx = \flowhum_\pathidx + \flowaut_\pathidx$ is a function of the density ($\denshum_\pathidx$ and $\densaut_\pathidx$, respectively) of each vehicle type as follows.
\begin{align}\label{eq:flow}
&\flow_\pathidx(\denshum_\pathidx, \densaut_\pathidx) := \nonumber \\
&\begin{cases}
\ffvelocity_\pathidx \cdot (\denshum_\pathidx + \densaut_\pathidx) ,& \text{if } \denshum_\pathidx + \densaut_\pathidx \le \critdens_{\pathidx}(\autlev_{\pathidx})\\
\frac{\ffvelocity_\pathidx \cdot \critdens_{\pathidx}(\autlev_{\pathidx}) \cdot (\jamden_\pathidx - (\denshum_\pathidx + \densaut_\pathidx)) }{\jamden_\pathidx - \critdens_{\pathidx}(\autlev_{\pathidx})} ,              & \text{if }  \critdens_{\pathidx}(\autlev_{\pathidx}) \le \denshum_\pathidx + \densaut_\pathidx \le \jamden_\pathidx \\
0 ,& \text{otherwise } \; .
\end{cases}
\end{align}

We can then write the latency as a function of vehicle flow as well as a binary argument $\congprof_\pathidx$, which indicates whether the road is congested \cite{krichene2017stackelberg,biyik2018altruistic}:
\begin{align}\label{eq:latency}
&\latency_\pathidx(\flowhum_\pathidx,\flowaut_\pathidx,\congprof_\pathidx) = \begin{cases}
\frac{d_\pathidx}{\ffvelocity_\pathidx} & \text{if } \congprof_\pathidx = 0 \\
d_\pathidx\left(\frac{\jamden_\pathidx}{\flowhum_\pathidx + \flowaut_\pathidx} + \frac{\critdens_\pathidx(\autlev_{\pathidx}) - \jamden_\pathidx}{\ffvelocity_\pathidx \cdot \critdens_\pathidx(\autlev_{\pathidx})}\right) & \text{if } \congprof_\pathidx = 1 \; .
\end{cases}
\end{align}
Fig.~\ref{fig:mixed_fund_diag_and_laten}(b) illustrates the effect of mixed autonomy on latency.

\subsection{Network Model}
\label{subsec:network_model}

\begin{assumption}\label{asmpt:ffl}
We consider a network of $\numpaths$ parallel roads. We assume that no two roads have the same free-flow latency. We order the indices such that $\fflatency_1 < \fflatency_2 < \ldots < \fflatency_\numpaths$.
\end{assumption}

The role of the assumption above is explained in the Appendix. We use $[k]$ to denote the set of the first $k$ roads; accordingly, $\pathset$ denotes the set of all roads.

We describe the network state by $(\flowhumvec, \flowautvec, \congprofvec)$, where $\flowhumvec$, $\flowautvec \in \mathbb{R}^\numpaths_{\ge 0}$ and $\congprofvec \in \{0,1\}^\numpaths$. A \emph{feasible routing} is one for which $\flowhum_\pathidx + \flowaut_\pathidx \le \capacity_\pathidx(\autlev_\pathidx)$ for all roads, and the flow of each vehicle type on the roads sum to $\demandhum$ and $\demandaut$ in the case of inelastic demand, respectively, which denote the total vehicle flow demands. In the case where the demand for autonomous vehicles is elastic, i.e., Section~\ref{sct:human_choice}, this constraint is relaxed such that the total autonomous flow does not exceed the maximum demand. We are interested in finding a routing, i.e. allocation of vehicles into the roads, that minimizes the total latency experienced by all vehicles while maximizing the total flow of the roads in the case of elastic demand.
Further, we constrain this optimization based on \emph{total demand}, \emph{selfishness} or \emph{flexibility} of the vehicles. While selfish drivers always take the quickest road possible, flexible vehicles accept relatively longer latencies. We will formalize these terms in Section~\ref{sct:equilibria}.

\subsection{Human Driver Choice Model}
\label{subsec:human_driver_choice}
\begin{figure}
	\centering
	\includegraphics[width=\columnwidth]{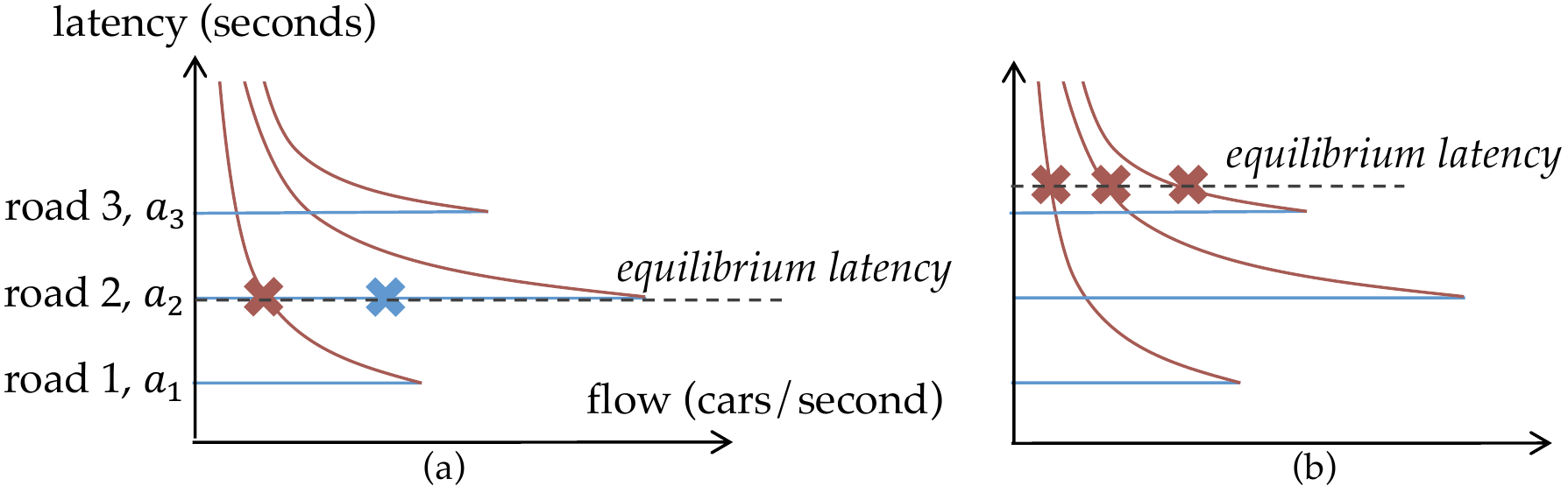}
	\vspace{-15px}
	\caption{Some possible equilibria of a three-road network with fixed flow demand. Blue and red lines denote the free-flow and congested regimes, respectively. Equilibria may have (a) one road in free-flow or (b) all used roads congested. An equilibrium has an associated \emph{equilibrium latency} experienced by all selfish users. By considering a given equilibrium latency, we can reason about which roads must be congested at that equilibrium as well how much flow is on each road.}
	\label{fig:equilibria}
	\vspace{-15px}
\end{figure}

We assume human drivers are selfish, i.e. their only consideration is minimizing their own commute time. This leads to a \emph{Nash Equilibrium} \cite{dafermos1980variationalinequalities}, which, on parallel roads, means that if one road has positive flow on it, all other roads must have higher or equal latency. Formally, 
$$\flowhum_\pathidx>0 \implies \latency_\pathidx(\flowhum_\pathidx, \flowaut_\pathidx, \congprof_\pathidx) \le \latency_{\pathidx'}(\flowhum_{\pathidx'}, \flowaut_{\pathidx'}, \congprof_{\pathidx'}) \; \forall \pathidx,\pathidx' \in \pathset \; .$$
This implies that all selfish users experience the same latency. It is therefore useful to consider the flow-latency diagrams of roads when studying which equilibria exist -- by fixing the latency on the y-axis, one can reason about which roads must be congested to achieve that equilibrium. As shown in Fig.~\ref{fig:equilibria}, equilibria may have one road in free-flow and rest congested, or all may be congested \cite{krichene2017stackelberg, biyik2018altruistic}.

\subsection{System Objective}\label{sct:optimization}
In this paper, we are interested in developing a pricing scheme for autonomous vehicles that improves the state of the traffic by alleviating the adverse effects of selfishness. First, we develop two baselines: the first is the case in which all users, including autonomous vehicles, are selfish and the network reaches a \emph{Nash Equilibrium}. In this case we wish to efficiently calculate the Nash Equilibrium that minimizes overall travel latency. Our goal is to achieve lower travel latencies than this equilibrium with the same amount of flow. In the second baseline, we assume we have limited direct control over the routing of autonomous users -- more specifically, we can route autonomous vehicles as we wish as long as the latency they experience is within some range of the quickest route available. A specific case of this baseline with full flexibility of the autonomous vehicles serves as a lower bound to our method.

After developing the two baselines in Section~\ref{sct:equilibria}, we present our pricing method that provides the same benefits as flexible behavior through financial incentives in Section~\ref{sct:human_choice}. For our pricing scheme, we describe the various facets of the problem. We assume the demand of human drivers is fixed, and the demand of people using the autonomous service is elastic -- \emph{if prices and latencies are high, some people may choose not to use the autonomous mobility service.} Our goal is then to simultaneously maximize the number of autonomous service users that can use the road, and minimize the average latency experienced by all the people using the roads.

\section{PERFORMANCE BENCHMARKS}\label{sct:equilibria}
Throughout this section, we assume an inelastic demand, i.e., we will route all autonomous and human-driven flow demand, $\demandaut$ and $\demandhum$, into the network. As the demand is inelastic, we are only interested in finding a routing that minimizes the total latency experienced by all vehicles, $C(\flowhumvec,\flowautvec,\congprofvec)=\sum_{\pathidx\in \pathset}(\flowhum_\pathidx + \flowaut_\pathidx)\latency_{\pathidx}(\flowhum_{\pathidx},\flowaut_{\pathidx},\congprof_\pathidx)$, while satisfying the demand, i.e. $\sum_{\pathidx\in \pathset}\flowhum_\pathidx = \demandhum$ and $\sum_{\pathidx\in \pathset}\flowaut_\pathidx = \demandaut$.

We now make precise the aforementioned notions of selfishness and flexibility. We develop properties of the resulting equilibria, and using those, provide polynomial-time algorithms for computing the benchmark flows.

\noindent\textbf{Selfishness.} Human drivers are often thought of as selfish, meaning they will not take a route with long latency if a quicker route is available to them. If all drivers are selfish this leads to a \emph{Nash Equilibrium}, in which no driver can achieve a lower travel time by unilaterally switching routes \cite{dafermos1980variationalinequalities}. This means that all selfish users with the same origin and destination experience the same travel time.
\begin{definition}\label{def:longest_eq_road}
	The \textbf{\emph{longest equilibrium road}} is the road with maximum free-flow latency which has latency equal to the latency experienced by selfish users. Let $\meq$ denote the index of this road. We use $\text{NE}(\demandhum,\demandaut,\meq)$ to denote the set of Nash Equilibria with \emph{longest equilibrium road} having index $\meq$. 
\end{definition} 
\begin{definition}\label{def:longest_used_road}
	The \textbf{\emph{longest used road}} is the road with maximum free-flow latency that has positive vehicle flow of any type on it. We use $\mall$ to denote the index of this road; if all vehicles in a network are selfish then $\meq=\mall$.
\end{definition}

The following lemma will help with the subsequent theoretical results; we defer its proof to the appendix.

\begin{lemma}\label{lma:free_flow}
	If the set of Nash Equilibria contains a routing with positive flow only on roads $\midcs$, then there exists a routing in the set of Nash Equilibria with positive flow only on roads $\mprimeidcs$ where $m' \le m$, and road $m'$ is in free-flow. 
\end{lemma}

We define the set of \textbf{\emph{Best-case Nash Equilibria (BNE)}} as the set of feasible routings in equilibrium that minimize the total latency for flow demand $(\demandhum, \demandaut)$, denoted  $\text{BNE}(\demandhum, \demandaut)$. The following theorem (proof deferred to the appendix) provides properties of the set of BNE for mixed-autonomy roads (for roads with a single vehicle type, see \cite{krichene2017stackelberg}).
\begin{theorem}\label{thm:BNE}
	There exists a road index $\mstareq$ such that all routings in the set of BNE have the below properties. Further, this index $\mstareq$ is the minimum index such that a feasible routing can satisfy the properties:
	\begin{enumerate}[nosep]
		\item road $\mstareq$ is in free-flow,
		\item roads with index less than $\mstareq$ are congested with latency $a_{\mstareq}$, and
		\item all roads with index greater than $\mstareq$ have zero flow.
	\end{enumerate}
\end{theorem}


As the same latency level can be achieved by varying the autonomy levels of the roads, BNE is not necessarily unique.

\noindent\textbf{Flexibility.} We also wish to find a lower bound for the social cost when some users are willing (or incentivized) to take longer routes. To that end, we use the term \emph{flexibility profile} to refer to the distribution of the degree to which autonomous users are willing to endure longer routes. For computational reasons, we consider flexibility profiles with a finite number of flexibility levels.

\begin{figure}
	\centering
	\includegraphics[width=\columnwidth]{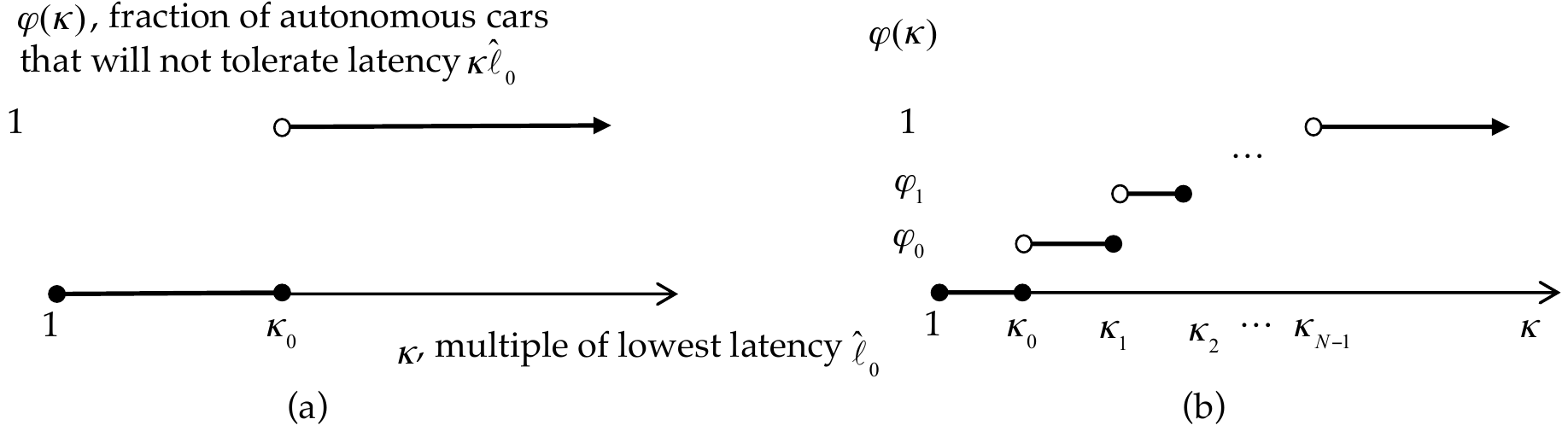}
	\vspace{-15px}
	\caption{Flexibility profiles. A fraction $\altruismFn(\altruismArg)$ of autonomous users will not accept latency greater than $\altruismArg$ times that of the quickest available route. \textbf{(a)} Users will tolerate latency of up to $\altruismUnif$ times that of the quickest route. \textbf{(b)} Users have multiple flexibility levels.} 
	\label{fig:altruism}
	\vspace{-15px}
\end{figure}

Formally, we define $\altruismFn: \mathbb{R}_{\ge 0} \rightarrow {[}0,1{]}$ to represent the flexibility profile as a nondecreasing function of a latency value that is mapped to ${[}0,1{]}$. A volume of $\altruismFn(\altruismArg)\demandaut$ autonomous flow will reject a route incurring latency $\altruismArg$ times the minimum route latency available, which we denote $\latency_\meq$. If autonomous users have a uniform flexibility level as in Fig.~\ref{fig:altruism}(a), we call them \emph{$\altruismUnif$-flexible} users, where $\altruismUnif$ is the maximum multiple of the minimum latency that autonomous users will accept. Users may have differing flexibility levels, as in Fig.~\ref{fig:altruism}(b). We use $\altruismLevSet$ to denote the set of flexibility levels, with cardinality $|\altruismLevSet|$. Accordingly, a feasible routing $(\flowhumvec, \flowautvec, \congprofvec)$ is in the set of \textbf{\emph{Flexible Nash Equilibria (FNE)}} if 
\begin{enumerate}[nosep]
	\item all routes with human traffic have latency $\latency_\meq \le \latency_\pathidx(\flowhum_\pathidx,\flowaut_\pathidx,\congprof_\pathidx)$ $\forall \pathidx \in \pathset$ and
	\item for any $
	\latency \geq 0$, a volume of at least $\altruismFn(\latency/\latency_\meq)\demandaut$ autonomous traffic experiences a latency less than or equal to $\latency$. Note that it is sufficient to check this condition for $\latency = \latency_\pathidx(\flowhum_\pathidx,\flowaut_\pathidx,\congprof_\pathidx)$ for all $\pathidx$.
\end{enumerate}

We denote the set of routings at Flexible Nash Equilibria with demand $(\demandhum,\demandaut)$, equilibrium latency $\latency_\meq$, and flexibility profile $\altruismFn$ as $\text{FNE}(\demandhum,\demandaut,\latency_\meq,\altruismFn)$. The set of \textbf{\emph{Best-case Flexible Nash Equilibria (BFNE)}} is the subset of FNE with routings that minimize total latency. Note that as in Theorem~\ref{thm:BNE}, we use $\mstareq$ to denote the road with longest free-flow latency that contains selfish vehicle flow in the best routing within the considered set of equilibria. We defer the proof of the following to the appendix.
\begin{theorem}\label{thm:BANE}
	For any given routing in the set of BFNE, there exist a longest equilibrium road $\mstareq$ and a longest used road $\mstarall$ with $\mstareq \le \mstarall$, such that:
	\begin{enumerate}[nosep]
		\item roads with index less than $\mstareq$ are congested,
		\item roads with index greater than $\mstareq$ are in free-flow,
		\item roads with index greater than $\mstareq$ and less than $\mstarall$ have maximum flow.
	\end{enumerate}
\end{theorem}


\begin{remark}
	Note that, unlike in BNE, road $\mstareq$ will not necessarily be in free-flow and $\mstareq$ is not necessarily the minimum index such that all selfish traffic can be feasibly routed at Nash Equilibrium \cite{biyik2018altruistic}. Further, different elements of the set BFNE can have different indices for longest equilibrium and longest used road.
\end{remark}

\noindent\textbf{Finding the Best-case Nash Equilibria.} In general, the Nash Equilibrium constraint is a difficult combinatorial constraint. Theorem \ref{thm:BNE} however states that we can characterize the congestion profile of the roads by finding the minimum free-flow road such that Nash Equilibrium can be feasibly achieved. This is formalized as follows: find the minimum $\meq$ such that $\text{NE}(\demandhum,\demandaut, \meq)$ is nonempty:
\begin{align}\label{opt:robust_BNE}
& \qquad \mstareq = \argmin_{\meq \in \pathset} a_\meq \; \text{s.t.} \; \text{NE}(\demandhum,\demandaut, \meq) \neq \emptyset \; \text{, then} \nonumber \\
&\text{BNE}(\demandhum,\demandaut) \; \subseteq \; \text{NE}(\demandhum,\demandaut, \mstareq) .
\end{align}
\begin{theorem}\label{thm:compute_BNE}
	\eqref{opt:robust_BNE} can be solved in $O(N^4)$ time.
\end{theorem}
We defer the proof to the appendix.

\noindent\textbf{Finding the Best-case Flexible Nash Equilibria.} To find an element of the BFNE, we need to solve:
\begin{align}\label{opt:BANE}
\argmin_{\substack{\meq \in \pathset, \; \; \eqDelay \in {[}a_\meq,a_{\meq+1}{)},\\ (\flowhumvec,\flowautvec,\bs) \in \text{FNE}(\demandhum,\demandaut,\eqDelay,\altruismFn)} } C(\flowhumvec,\flowautvec,\congprofvec) \; .
\end{align}
As demonstrated in \cite{biyik2018altruistic}, the longest equilibrium road is no longer the road with lowest free-flow latency such that the routing is feasible, as was the case in BNE. Further, road $\meq$ may not be in free-flow in the set of BFNE. However, we do know that for a fixed $\meq$, the latency on road $\meq$ which minimizes cost, subject to feasibility constraints, is one of a finite number of options.

\begin{theorem}\label{thm:compute_BANE}
	Finding a solution to \eqref{opt:BANE} is equivalent to finding a routing in the set of BFNE, if any exist. Further, \eqref{opt:BANE} can be solved in $O(|\altruismLevSet|N^5)$ time, where $|\altruismLevSet|$ is the number of flexibility levels of autonomous vehicle users.
\end{theorem}

\begin{proof} 

First, Definition~\ref{def:longest_eq_road} implies that the latency on the longest equilibrium road $\meq$, which we denote $\eqDelay$, must be less than that of road $\mstareq+1$. This, with the definition of BFNE imply that \eqref{opt:BANE} solves for an element of the BFNE. Now, note that for a given $\eqDelay$, the optimal routing will maximize the autonomous flow on roads $\meqidcs$. We show that this can be computed in $O(N^3)$ time, and the optimal allocation of the remaining autonomous flow can be computed in $O(N)$ time. Next, we note that subject to feasibility, the social cost of a routing decreases monotonically with $\eqDelay$. In light of this, we show that there are a maximum of $k|N|$ critical points to check for feasibility when searching for the optimal $\eqDelay$.

We now show that given $\eqDelay$, the latency on road $\meq$, computing the optimal flow on roads $\meqidcs$ can be done in $O(N^3)$ time. For a given $\eqDelay$, the optimal routing will fit as much autonomous flow as possible on roads $\meqidcs$. Let $\flowhumvecEQ$ and $\flowautvecEQ$ denote the elements of the regular and autonomous routings $\flowhumvec$ and $\flowautvec$ that correspond to flows on the roads $\meqidcs$, as with $\congprofvec^\text{EQ}$. Then, solve
\begin{align}\label{opt:max_aut_on_meq_roads}
\argmax_{\flowhumvecEQ, \flowautvecEQ \in \mathbb{R}^{|\meqidcs|}_{\ge 0},\congprof_\meq \in \{0,1\} }\;
& \sum_{\pathidx\in \meqidcs}\flowaut_i \\
\text{s.t.}\;
& \sum_{\pathidx\in \meqidcs}\!\flowhum_\pathidx \!=\! \demandhum, \sum_{\pathidx\in \meqidcs}\!\flowaut_\pathidx \!\leq\! \demandaut, \nonumber\\
\quad \forall \pathidx \in \meqidcs\;
& \flowhum_\pathidx \ge 0, \flowaut_\pathidx \ge 0,  \nonumber\\
\quad \forall i \in {[}\meq-1{]} \;
& \latency_\pathidx(\flowhum_\pathidx,\flowaut_\pathidx,1)=\eqDelay, \nonumber\\
& \latency_\meq(\flowhum_\meq,\!\flowaut_\meq,\!\congprof_\meq)\!=\!\eqDelay .\nonumber
\end{align}

Using similar reasoning as in Theorem~\ref{thm:compute_BNE}, this can be formulated as a linear program and therefore can be solved with a computational complexity $O(N^3)$ \cite{gonzaga1992path}.

Having computed the optimal routing on roads $\meqidcs$, we now consider an optimization which computes the resulting optimal index of the longest used road in order to fit the autonomous flow, and as a result, the optimal routing of the remaining autonomous flow. This follows from Theorem~\ref{thm:BANE}:
\begin{equation}\label{opt:num_ff_roads_BANE}
\begin{aligned}
& \argmin_{j\in\pathset \backslash {[}\meq-1{]}}
& & \quad j \\
& \qquad \quad \text{s.t.}
& & \sum_{\pathidx\in \meqidcs}\flowaut_\pathidx + \sum_{\pathidx \in {[}j{]} \backslash \meqidcs}\capacity_\pathidx(1)\ge \demandaut \; ,
\end{aligned}
\end{equation}
which requires computations of order $O(N)$.

We temporarily restrict our attention to the case in which autonomous users have a uniform autonomy level. We wish to optimize over the following decision variables:
\begin{equation*}
\begin{aligned}
& \meq \in \pathset
& & \text{longest equilibrium road?} \\
& \mall \in \midcs \backslash {[}\meq-1{]} 
& & \text{longest used road?} \\
& \eqDelay \in {[}\fflatency_\meq,\fflatency_{\meq+1}{)} 
& & \text{equilibrium latency?} \\
& \flowhumvec,\flowautvec \in \mathbb{R}^n_{\ge 0}, \bs \in \{0,1\}^n 
& & \text{actual routing?}
\end{aligned}
\end{equation*}

The objective function to be minimized is aggregate latency, which, using the Theorem~\ref{thm:BANE}, can be formulated as follows:
\begin{align}
& \eqDelay \sum_{\mathclap{\pathidx \in \meqidcs}}(\flowhum_\pathidx+\flowaut_\pathidx) + \sum_{\mathclap{\pathidx\in {[}\mall-1{]} \backslash \meqidcs}}a_\pathidx \capacity_\pathidx(1) + a_\mall\Big(\demandaut-\sum_{\mathclap{\pathidx \in {[}\mall-1{]}}}\flowaut_\pathidx\Big) \nonumber \\ 
\text{s.t. } & (\flowhumvecEQ, \flowautvecEQ, \bs^{\text{EQ}}) \in \eqref{opt:max_aut_on_meq_roads} \label{eq:max_aut_flow} \\
& \mall = \eqref{opt:num_ff_roads_BANE} \label{eq:feas_longest_road} \\
& \frac{1}{\demandaut} \Big( \sum_{\mathclap{\pathidx \in \meqidcs}}\flowaut_\pathidx + \sum_{\mathclap{\pathidx \in {[}j{]} \backslash \meqidcs}}\capacity_\pathidx(1) \Big) \ge \altruismFn(\frac{a_i}{\eqDelay}) \; \forall j \in \pathset \backslash {[}\meq-1{]} \label{eq:altruism}
\end{align}

For a given $\eqDelay$, the optimal routing maximizes the autonomous flow on roads $\meqidcs$, yielding \eqref{eq:max_aut_flow}. The remaining autonomous flow is routed as in \eqref{eq:feas_longest_road}. Finally, \eqref{eq:altruism} ensures that no one is more flexible than they wish.

To solve this, recall that in the case of uniform flexibility,
\begin{align*}
\altruismFn(\altruismArg) = \begin{cases}
0 & 0 \le \altruismArg \le \altruismUnif \\
1 & \altruismArg > \altruismUnif \; .
\end{cases}
\end{align*}

Further, the volume of autonomous flow that can fit on roads $\midcs$ increases with decreasing $\latency_m$. Because of this, we can restrict our search of $\eqDelay$ to critical points of the function $\altruismFn$:
\begin{align*}
\eqDelay \in \{\fflatency_\meq\}\! \cup\! \{\frac{\fflatency_\pathidx}{\altruismUnif}\! :\! \pathidx\!\in\! \mallidcs \backslash {[}\meq{]}, \fflatency_\meq\! <\! \frac{\fflatency_\pathidx}{\altruismUnif}\! <\! \fflatency_{\meq+1} \},
\end{align*}
which is a set with maximum cardinality $N$. Therefore, we can find the BFNE via the following algorithm:
\begin{enumerate}
	\item Enumerate through all possible values of $\meq$ ($N$ possibilities).
	\item For each possible $\meq$, enumerate through all possible values of $\eqDelay$ ($N$ possibilities).
	\item For each combination of $\meq$ and $\eqDelay$, find the optimal routing on roads $\meqidcs$ via \eqref{opt:max_aut_on_meq_roads} (order $N^3$), and find $\mall$ and the optimal routing of autonomous vehicles on the remaining roads via \eqref{opt:num_ff_roads_BANE} (order $N$). As these are sequential, this step requires computations of order $O(N^3)$.
\end{enumerate}
All together, this requires computations of order $O(N^5)$.

Now consider that autonomous vehicles have nonuniform flexibility levels. Then,
\begin{align*}
\altruismFn(\altruismArg) = \begin{cases}
0 & 0 \le \altruismArg \le \altruismArg_0 \\
\varphi_0 & \altruismArg_0 < \altruismArg \le \altruismArg_1 \\
\varphi_1 & \altruismArg_1 < \altruismArg \le \altruismArg_2 \\
\ldots & \\
1 & \altruismArg > \altruismArg_{{|\altruismLevSet|-1}} \; .
\end{cases}
\end{align*}
We therefore must search over $\eqDelay$ in the following set:
\begin{align*}
&\eqDelay\! \in\{\fflatency_\meq\} \cup \\
& \{\frac{\fflatency_i}{\altruismArg_j}\! :\! \pathidx\!\in\! \mallidcs \backslash {[}\meq{]}, j\! \in\! \altruismLevSet, \fflatency_\meq\! <\! \frac{\fflatency_\pathidx}{\altruismArg_j}\! <\! \fflatency_{\meq+1} \} ,
\end{align*}
which has maximum cardinality $|\altruismLevSet|N$, bringing the total computation complexity to $O(|\altruismLevSet|N^5)$.
\end{proof}

\section{INCENTIVIZING FLEXIBILITY}\label{sct:human_choice}
We next present a pricing mechanism to attain the same benefits of flexible behavior through financial incentives in the case of elastic demands. We first explain the human choice model that formalizes how autonomous service users choose between a variety of price and latency pairs depending on their value of time, encoded by their \emph{reward functions}. We then formulate a mathematical optimization problem that aims to find an optimal trade-off between high road usage and low average travel time. We then explain the data-efficient algorithm that we propose for learning human reward functions.

\subsection{Autonomous Service User Choice Model}
\label{subsec:human_choice}
How users choose between a variety of price and latency pairs depends on their valuation of time and money. Without knowing this choice model we cannot plan vehicle flows and will not be able to ensure the resulting configuration matches our vehicle flow models for the roads. Also, since different populations may have different valuations, we need to learn this tradeoff for our population so we can estimate how many people will choose which option.

To untangle these constraints, we describe how autonomous service users make routing decisions (the model for human drivers is provided in Sec.~\ref{subsec:human_driver_choice}). Though human drivers are motivated directly only by latency, autonomous service users experience cost in both latency and the price of the ride. We model the users as having some underlying reward function, which is parameterized by their time/money tradeoff, as well as the desirability of the option of traveling by some other means such as walking or public transit. We assume that a strictly dominated option in terms of latency and price is completely undesirable. We formally define this set
\begin{align*}
D = \{\pathidx \in \pathset \: \mid\:(\price_\pathidx>\price_{\pathidx'} \land \latency_\pathidx \geq \latency_{\pathidx'}) \lor (\price_\pathidx \geq \price_{\pathidx'} \land \latency_\pathidx > \latency_{\pathidx'})\\ \textrm{ for some } \pathidx'\in \pathset\} \; ,
\end{align*}
where $\price_\pathidx$ is the monetary cost of route $\pathidx$. We also define the set of undominated roads, $\nondominatedroads = \pathset \backslash \dominatedroads$. We model the reward function of user $\useridx$ for choosing road $\pathidx$ as follows, where we assume choosing road $0$ denotes declining the service:
\begin{align*}
&\reward_\useridx(\boldsymbol{\latency},\mathbf{\price},\pathidx) = \\
\quad &\begin{cases}
- {\omega_\useridx}_1 \latency_\pathidx - {\omega_\useridx}_2 \price_\pathidx & \textrm{if }\pathidx \in \nondominatedroads, \\
-\infty & \textrm{if }\pathidx \in \dominatedroads,\\
-\zeta_\useridx \latency^w & \textrm{if }\pathidx\!=\!0\textrm{ (user }\useridx \textrm{ declines the service) ,}
\end{cases}
\end{align*}
with the following nonnegative parameters: $\latencyvec$ denotes the vector of road latencies, $\pricevec$ denotes prices, ${\boldsymbol{\omega}_\useridx} = \begin{bmatrix}{\omega_\useridx}_1 & {\omega_\useridx}_2 \end{bmatrix}^\top$ characterizes the users' time/money tradeoff and $\zeta_\useridx$ specifies their willingness to use an alternative option with latency $\latency^w$, which could be walking, biking, or public transportation.

We do not assume users are simple reward maximizers. Rather, we adopt the multinomial logit model \cite{ben1985discrete} for the probability with which users choose each option:
\begin{equation}\label{eq:decision_probs}
\!P(\textrm{user }\useridx \textrm{ chooses route } \pathidx) \propto \exp\left(\reward_\useridx(\latencyvec,\pricevec,\pathidx)\right)
\end{equation}
for all $\pathidx\in\pathset\cup\{0\}$.

In order to determine the optimal pricing, we want to know how many users will choose each route as a function of the route prices and latencies. To this end, we define $q_i(\latencyvec, \pricevec)$ as the expected fraction of autonomous service users that will choose route $\pathidx$. If the parameter distribution for autonomous service users is $g(\boldsymbol{\omega},\zeta)$, then
\begin{align*}
&q_i(\latencyvec, \pricevec) = \int_0^\infty\int_0^\infty\int_0^\infty g(\boldsymbol{\omega},\zeta)P(\pathidx \mid \boldsymbol{\omega}, \zeta) d{\omega_1} d{\omega_2} d\zeta \; .
\end{align*}
where $P(i \mid \boldsymbol{\omega}, \zeta)$ is the probability that a user with reward function parameters $(\boldsymbol{\omega},\zeta)$ will choose road $\pathidx$. This expression relate prices and latency to human choices, enabling us to determine the prices that will maximize the social objective. This will be important in constraining the optimization to only consider latency/price options that correspond to the desired vehicle flows.

\subsection{Solution Method}

\noindent \textbf{Problem Formulation. } We now formulate the problem where we have an indirect control of the autonomous cars' routing through pricing, and the demand of autonomous service is elastic. Because of this elasticity, we cannot just minimize the average latency, which would result in extremely high prices to keep autonomous service users off the network. Hence, we consider an objective that is a combination of maximizing road usage and minimizing average travel time.\footnote{Some other works, e.g. \cite{chau2003price}, use the \emph{social welfare} objective. While the two objectives have many similarities, we cannot directly adopt it here as we have heterogeneous users who have different price-latency valuations.} We parameterize this tradeoff with parameter $\theta \ge 0$ in the cost function
\begin{equation*}
J(\flowhumvec,\!\flowautvec,\!\congprofvec) \!=\! \frac{\sum_{\pathidx \in \pathset}{(\flowhum_\pathidx\!+\!\flowaut_\pathidx) \latency_\pathidx(\flowhum_\pathidx, \flowaut_\pathidx, \congprof_\pathidx)}}{\sum_{\pathidx \in \pathset}{(\flowhum_\pathidx+\flowaut_\pathidx)}} - \theta\!\sum_{\pathidx \in \pathset}\! (\flowhum_\pathidx\!+\!\flowaut_\pathidx) \; .
\end{equation*}

Our control variables in this optimization are the \emph{latency} on each road and the \emph{price} offered to the users for traveling on each route. However, we cannot arbitrarily choose prices and latencies -- we need to respect 
\begin{enumerate}
	\item \emph{the characteristics of the roads}, in terms of how the flow demand for a road corresponds to the latency on that road (Section~\ref{subsec:vehicle_flow}), and 
	\item \emph{how people make decisions}, making sure that the number of people who choose each option corresponds to the latency of the roads described in the options (Section~\ref{subsec:human_choice}).
\end{enumerate}
Moreover, we want to be fair; so we must offer the same pricing and routing options to all autonomous service users.

Given $\mathbf{q}(\latencyvec,\pricevec)$, the cost function $J(\flowhumvec, \flowautvec, \congprofvec)$, inelastic flow demand of human drivers $\demandhum$, and elastic demand of autonomous users $\demandaut$, we are ready to formulate the planning optimization. The most straightforward way is to optimize jointly over $\flowhumvec,\flowautvec,\latencyvec$ and $\bp$. However, $\boldell$ is fully defined by $\flowhumvec,\flowautvec$ and $\congprofvec$. Hence, instead of $\latencyvec$, we use $\congprofvec$, which will help us solve this nonconvex optimization. The problem is then formulated as:
\begin{flalign}
\min_{\flowhumvec,\flowautvec,\pricevec \in \mathbb{R}^\numpaths_{\ge 0},k \in \pathset,\congprof_{k:N}\in\{0,1\}^{N-k+1}}  J(\flowhumvec,\flowautvec,\congprofvec)&&
\label{eq:optimization}
\end{flalign}
\vspace{-10px}
\begin{align}
\textrm{s.t.} & \sum_{\pathidx \in \pathset} \flowhum_\pathidx = \demandhum \label{c:1}\\
& \flowaut_\pathidx = \demandaut q_i(\latencyvec(\flowhumvec,\flowautvec,\congprofvec), \pricevec), \forall \pathidx \in \left(\pathset\setminus [k]\right) \cup D \label{c:2}\\
& \sum_{i \in [k] \setminus D}\flowaut_\pathidx = \demandaut \sum_{i \in [k] \setminus D}q_i(\latencyvec(\flowhumvec,\flowautvec,\congprofvec), \pricevec) \label{c:2_extra}\\
& a_k \leq \latency_k(\flowhum_k,\flowaut_k,\congprof_k) \leq a_{k+1} \label{c:3}\\
& \flowhum_\pathidx=0, \forall \pathidx \in \pathset \setminus {[}k{]} \label{c:4}\\
& \latency_\pathidx(\flowhum_\pathidx,\flowaut_\pathidx,1) = \latency_k(\flowhum_k,\flowaut_k,\congprof_k), \forall \pathidx \in {[}k-1{]} \label{c:5}\\
& \flowhum_\pathidx + \flowaut_\pathidx \leq \capacity\left(\flowaut_\pathidx/(\flowaut_\pathidx+\flowhum_\pathidx)\right), \forall \pathidx \in \pathset \label{c:6}\\
& \sum_{\pathidx \in \pathset} \left(\flowaut_\pathidx p_\pathidx - \flowaut_\pathidx d_\pathidx c\right) \geq \bar{P} \label{c:7}
\end{align}
with $\congprof_{1:k-1}=1$ due to selfishness of human-driven vehicles. Here, $k$ is the longest road with human-driven vehicle flow, $\bar{P}$ is the minimum profit per unit time for the autonomous service provider and $c$ is the constant fuel cost per unit length. We can describe the constraints as follows.
\begin{enumerate}[nosep]
	\setcounter{enumi}{12}
	\item The human-driven vehicle flow demand is fulfilled.
	\item Autonomous flow will be distributed into the roads in $\left(\pathset\setminus[k]\right)\cup D$ based on the choice model described in the preceding section.
	\item Total autonomous flow in $[k]\setminus D$ will satisfy the user choices, but can be distributed arbitrarily as the roads have the same latency and price.
	\item The ``longest equilibrium road" has latency on the given interval of free flow latency.
	\item Human-driven cars are selfish, i.e. no human-driven car will experience higher latency than the road $k$.
	\item The congested roads have the same latency as the ``longest equilibrium road".
	\item The maximum capacities of the roads are respected.
	\item The minimum profit per unit time is satisfied.
\end{enumerate}

We can further improve the search space by relying on the heuristic that the roads that are not used by the human-driven vehicles will be in free-flow, i.e. $s_{k+1:N}=0$. While we do not have a proof for the conditions that lead to this, we also note constructing counterexamples seems to require extremely careful design, which suggests the heuristic holds in general. Furthermore, the following theorem shows we could also set $s_k=0$ under an additional assumption.
\begin{theorem}\label{thm:human_choice}
	Assume $\omega_2>0$ for all users. Then there exists a free-flow road $k$ in the optimal solution to the problem such that $\latency_{\pathidx}=\latency_k$ for $\forall \pathidx\in[k]$, and $\flowhum_{\pathidx} = 0$ for $\forall \pathidx \in\pathset\setminus[k]$ as long as the optimization is feasible.
	\label{prop:eq_road}
\end{theorem}
We defer the proof to the appendix.

\noindent\textbf{Generalizations.} We assumed all autonomous cars are controlled by a centralized social planner. To extend our framework to scenarios where this is not the case and the social planner has the control over a fraction of autonomous cars, we can simply do the following modifications: The optimization will also be over $\flowautnocontrolvec\in\mathbb{R}_{\geq0}^\numpaths$, which will now represent the autonomous flow that does \emph{not} belong to the planner. Add the corresponding constraint of \eqref{c:1} for $\flowautnocontrol_\pathidx$. Similar to \eqref{c:4}, $\flowautnocontrol_\pathidx=0$ for $\pathidx \in \pathset \setminus {[}k{]}$ due to selfishness. Also, replace $\flowaut$'s in \eqref{c:3}, \eqref{c:5} and \eqref{c:6} with $\flowaut + \flowautnocontrol$ with appropriate subscripts. These simple modifications enable a more general use.

\noindent \textbf{Solving the Optimization. }
After learning the distribution $g(\boldsymbol{\omega},\zeta)$ (described below), we first take $M$ samples $(\boldsymbol{\bar\omega},\bar\zeta)\sim g(\boldsymbol{\omega},\zeta)$. Using these samples, we approximate the expected fraction of autonomous users that will choose route $\pathidx$ as:
\begin{align*}
&q_i(\latencyvec, \pricevec) \asymeq \frac{1}{M}\sum_{\boldsymbol{\bar\omega},\bar\zeta} P(\pathidx \mid \boldsymbol{\bar\omega}, \bar\zeta)
\end{align*}
where $\asymeq$ denotes asymptotic equality as $M\to\infty$.

We then locally solve the nonconvex planning optimization using interior point algorithms \cite{waltz2006interior} with $100$ random initial points for each run to get closer to global optimum.

\noindent \textbf{Data-Efficient Learning of Human Reward Functions. }
While the routes in a specific network can be fully modeled with the physical properties and the speed limits, user's decision models (parameterized by $(\boldsymbol{\omega}, \zeta)$) must be learned in a data-driven way. The parameters might be different among the users. While a business executive might prefer paying extra to reach their company quickly, a graduate student may decide to go to the lab a little later in order to save a few dollars. Therefore, we have to learn personalized parameters $\boldsymbol{\omega}$ and $\zeta$.

We learn the parameters from users' previous choices, which is known as \emph{preference-based learning}. If user $\useridx$ chooses from a variety of options, the user's choice gives us a noisy estimate of which road $\pathidx \in \pathset\cup\{0\}$ maximizes their reward function $\reward_\useridx(\latencyvec, \pricevec, \pathidx)$. We could start from either uniform priors or priors informed by domain knowledge, then sample from the distribution $g(\boldsymbol{\omega},\zeta)$.

However, a major drawback of doing so is how quickly we learn the user preferences. Preference-based learning suffers from the small amount of information that each query carries. For example, if we show $4$ options to a user (including the option to decline the service), then the maximum information we can get from that query is only $2$ bits. To tackle this issue, previous works pose the query synthesis problem as an optimization and maximize a measure of the expected change in the learned distribution after the queries \cite{sadigh2017active}. 


While those works focus on pairwise queries, in this case we expect to pose several route options to the users and therefore need more general query types. By using these general queries which offer a variety of routes with varied latency and price, we can consider various ways of using this learning framework to learn the human preferences.
\begin{itemize}
	\item We could do a user study on a few people to learn a good prior distribution.
	\item We could use an exploration/exploitation strategy if we are allowed to break the fairness constraint a few times for some small portion of the users. This could be made through user-specific promotions; for each user we may either choose to use the learned model or to offer special rates that would help us profile the user better.
	\item We could do an initial profiling study for each new user.
\end{itemize}

To implement any of these options, we formulate the following active learning optimization. First, we discuss the general preference-based learning framework. Given the data from previous choices of user $\useridx$, which we denote as $\dataset_\useridx$, we formalize the probability of $(\boldsymbol{\omega}_j,\zeta_j)$ being true parameters for that user as follows:
\begin{align*}
P(\boldsymbol{\omega}_\useridx,\zeta_\useridx \mid \dataset_\useridx) &\propto  P(\boldsymbol{\omega}_\useridx,\zeta_\useridx)\prod_{m}P({\dataset_\useridx}_m \mid \boldsymbol{\omega}_\useridx,\zeta_\useridx)
\end{align*}
where ${\dataset_\useridx}_m$ denotes the road user $\useridx$ chose in their $m^{\textrm{th}}$ choice (with ${\dataset_\useridx}_m=0$ meaning that the user declined the service and preferred the alternative option). The relation is due to the assumption that the users' choices are conditionally independent from each other given the reward function parameters.

The second term comes from the human choice model. For the prior, we can use a uniform distribution over nonnegative parameters. The prior may be crucial especially when we do not have enough data for a new user. In such settings, we incorporate domain knowledge to start with a better prior.

We then use this unnormalized $P(\boldsymbol{\omega}_\useridx,\zeta_\useridx \mid \dataset_\useridx)$ to obtain the samples of $(\boldsymbol{\omega}_\useridx,\zeta_\useridx)$ using Metropolis-Hastings algorithm. Doing this for each user, which can be easily parallelized, we directly obtain the samples $(\boldsymbol{\bar\omega},\bar\zeta)\sim g(\boldsymbol{\omega},\zeta)$.

Next we formulate the active learning framework, which is needed so that it will not take an excessive number of queries to learn human preferences. For this, we want to maximize the expectation of the difference between the prior and the unnormalized posterior:
\begin{align*}
\textrm{query}_m^* &= \argmax_{\textrm{query}_m} \mathbb{E}_{{\dataset_\useridx}_m}\Big[P(\boldsymbol{\omega}_\useridx,\zeta_\useridx|{\dataset_\useridx}_{1:m-1})-\\
&\qquad P(\boldsymbol{\omega}_\useridx,\zeta_\useridx|{\dataset_\useridx}_{1:m-1})P({\dataset_\useridx}_m | \boldsymbol{\omega}_\useridx,\zeta_\useridx,{\dataset_\useridx}_{1:m-1})\Big]\\
&= \argmin_{\textrm{query}_m} \mathbb{E}_{{\dataset_\useridx}_m}\left[P({\dataset_\useridx}_m | \boldsymbol{\omega}_\useridx,\zeta_\useridx,{\dataset_\useridx}_{1:m-1})\right]
\end{align*}
As we will use the sampled $(\boldsymbol{\omega}_\useridx,\zeta_\useridx)$ to compute the probabilities of road choices, we can write the optimization as:
\begin{align*}
\textrm{query}_m^* &\asymeq \argmin_{\textrm{query}_m} \mathbb{E}_{{\dataset_\useridx}_m}\!\left[\sum_{\boldsymbol{\bar\omega}_\useridx,\bar\zeta_\useridx}\!P({\dataset_\useridx}_m | \boldsymbol{\bar\omega}_\useridx,\bar\zeta_\useridx,{\dataset_\useridx}_{1:m\!-\!1})\right]
\end{align*}
where we have $M$ samples denoted as $(\boldsymbol{\bar\omega}_\useridx,\bar\zeta_\useridx)$, and the term $1/M$ is canceled. Using the law of total probability,
\begin{align*}
P({\dataset\!_\useridx}_m& | {\dataset\!_\useridx}_{1:m-1}) \asymeq\frac1M\sum_{\boldsymbol{\bar\omega}_\useridx,\bar\zeta_\useridx}P({\dataset_\useridx}_m | \boldsymbol{\bar\omega}_\useridx,\bar\zeta_\useridx,{\dataset_\useridx}_{1:m-1})
\end{align*}
which leads to the following optimization for finding $\textrm{query}_m^*$:
\begin{align*}
\argmin_{\textrm{query}_m}\! \sum_{{\dataset\!_\useridx}_m}P({\dataset\!_\useridx}_m|{\dataset\!_\useridx}_{1:m\!-\!1})\!\sum_{\boldsymbol{\bar\omega}_\useridx,\bar\zeta_\useridx}\!P({\dataset\!_\useridx}_m | \boldsymbol{\bar\omega}_\useridx,\bar\zeta_\useridx,{\dataset\!_\useridx}_{1:m\!-\!1})\\
\!\asymeq\!\argmin_{\textrm{query}_m}\! \sum_{{\dataset_\useridx}_m}\left(\sum_{\boldsymbol{\bar\omega}_\useridx,\bar\zeta_\useridx}P({\dataset_\useridx}_m | \boldsymbol{\bar\omega}_\useridx,\bar\zeta_\useridx)\right)^2
\end{align*}
We can easily compute this objective value for any given $\textrm{query}_m$. This optimization is nonconvex due to the human choice model. As in previous works, we assume local optima is good enough \cite{sadigh2017active}. We then use a Quasi-Newton method (L-BFGS \cite{andrew2007scalable}) to find the local optima, and we repeat this for $1000$ times starting from random initial points.

\section{USER EXPERIMENTS}\label{sct:user_experiments}
To validate our framework, we conducted different simulations and a user study approved by Stanford University's Research Compliance Office.

\noindent\textbf{Hypotheses.} We test three hypotheses that together suggest our framework successfully reduces traffic congestion through pricing, after it learns the humans' choice models:\\
\indent\textbf{H1}: Our active learning algorithm can learn the autonomous service user preferences in a data-efficient way.\\
\indent\textbf{H2}: Our planning optimization reduces the overall latency by creating flexible behavior through pricing.\\
\indent\textbf{H3}: When used by humans, the overall framework works well and is advantageous over inflexible algorithms.

\noindent\textbf{Implementation Details.} In the planning optimization, we used the heuristic $s_{k+1:N}\!=\!0$. We assumed the only alternative for autonomous service users is walking. We set $c\!=\!6\!\times\! 10^{-5}$ USD/meter, $\mathbf{b}\!=\!1$. We assumed the regular and autonomous cars keep a $2$-second and $1$-second headway distance with the leading car, respectively. The length of the cars is $5$ meters, and the minimum gap between two cars is $2$ meters.

\begin{figure}[h]
	\centering
	\includegraphics[width=0.85\columnwidth]{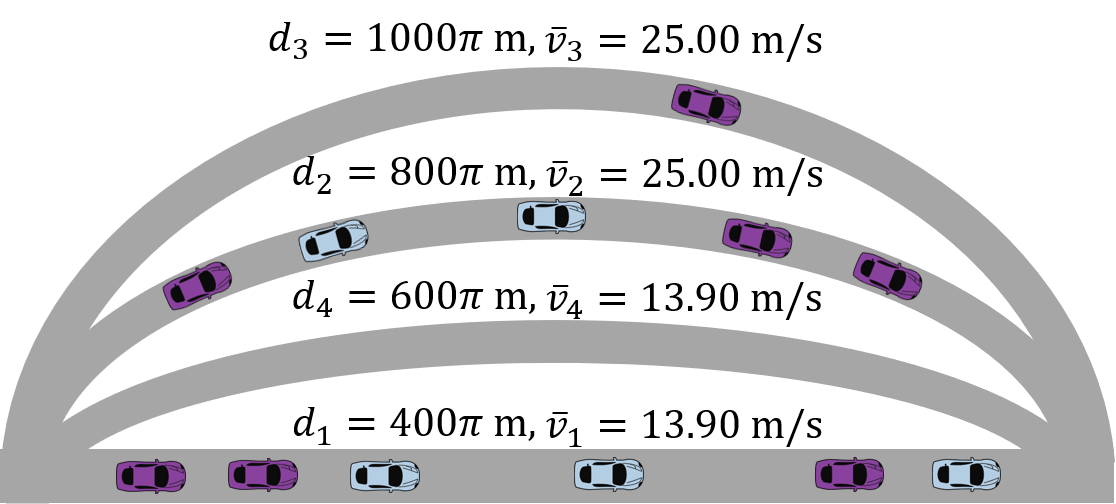}
	\caption{The $4$-road network from \cite{biyik2018altruistic}. The roads are not to the scale and ordered with respect to the free-flow latencies.}
	\label{fig:wafr_network}
	\vspace{-16px}
\end{figure}

\noindent\textbf{Experiments and Analyses.} To validate \textbf{H1}, we simulated $5$ autonomous service users with different preferences. We tested our active learning framework by asking two sets of $200$ queries, each of which consisted of $4$ road options, similar to Fig.~\ref{fig:wafr_network}, and a walking option. The queries were generated actively in the first set and randomly in the second. After each query, we recorded the sample $(\bar{\boldsymbol{\omega}},\bar\zeta)$ which has the highest likelihood as our estimates.

\begin{figure}[h]
	\centering
	\vspace{-10px}
	\includegraphics[width=\columnwidth]{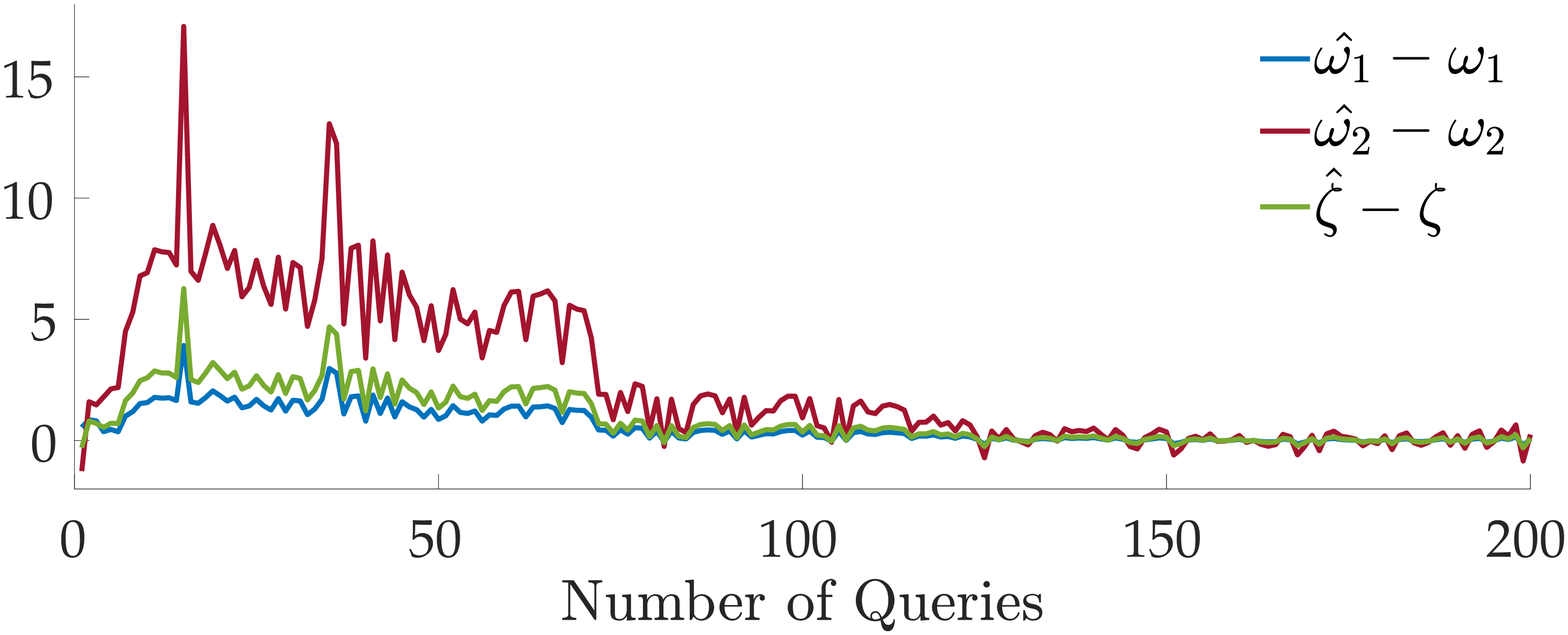}
	\vspace{-15px}
	\caption{The errors of the reward function estimates are shown with varying number of queries. $\hat{\omega_1}$, $\hat{\omega_2}$ and $\hat{\zeta}$ represent the estimates.}
	\label{fig:param_values}
	\vspace{-10px}
\end{figure}

Fig.~\ref{fig:param_values} shows how the estimates evolved within active learning setting for one of the users. All values are overestimated initially. Intuitively, this is because getting noiseless responses has higher likelihood in the beginning. As we query more, accepting some of the responses as noisy maximizes the likelihood. Therefore, the values start decreasing.

\begin{figure}[h]
	\centering
	\vspace{-10px}
	\includegraphics[width=\columnwidth]{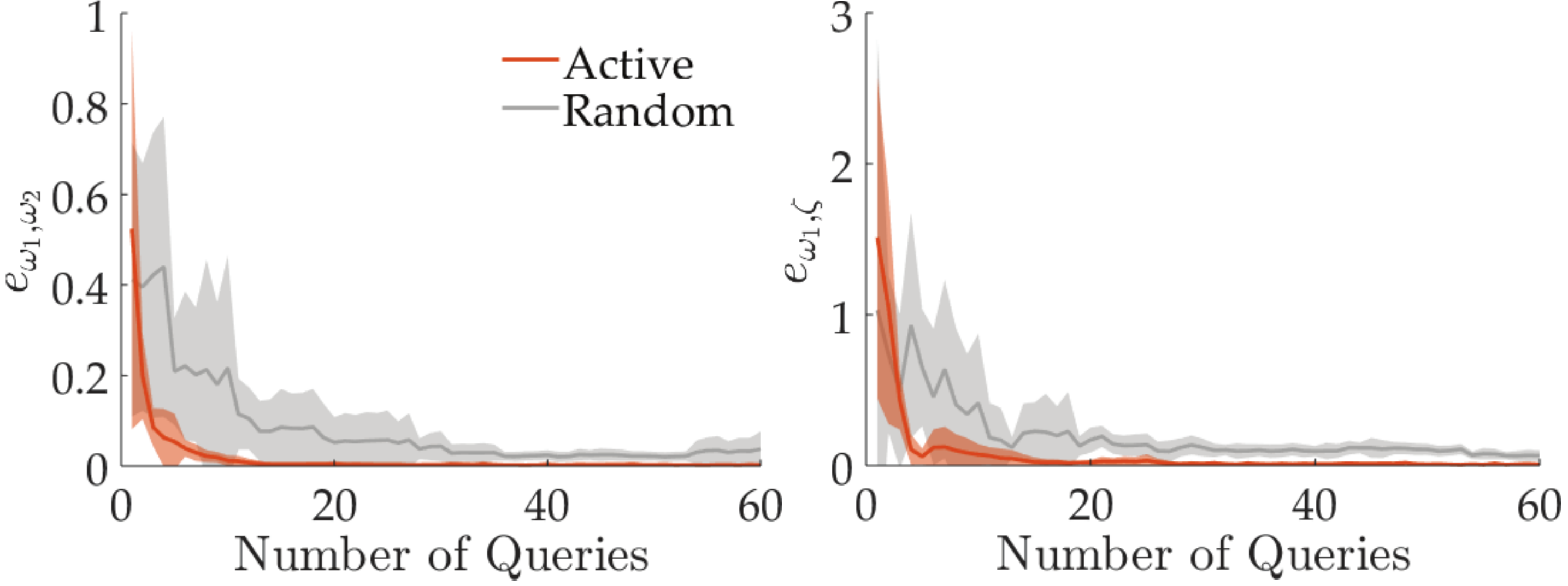}
	\vspace{-15px}
	\caption{The error metric is averaged over $5$ different reward functions.}
	\label{fig:active_vs_random}
	\vspace{-10px}
\end{figure}

Another important observation is that the estimates of the parameters increase and decrease together even in the early iterations. This suggests we are able to learn the ratio between the parameters, e.g. $\omega_1/\omega_2$ very quickly. To check this claim, we used the following error metric:
\begin{align*}
e_{x,y} = \norm{x/y - \hat{x}/\hat{y}}_1
\end{align*}
where $x,y\in\{\omega_1,\omega_2,\zeta\}$ and $\hat{x},\hat{y}$ represent the corresponding estimates. Fig.~\ref{fig:active_vs_random} shows how this error decreases with increasing number of queries. It also shows how active querying enables data-efficient learning compared to the random querying baseline. We are able to learn the relationship between parameters even under $20$ queries with active learning. All these results strongly support \textbf{H1}. 

The fact that we are able to learn the ratios implies we can estimate which road the user is most likely to choose. We will only be unsure about how noisy the user is if the parameter estimates did not converge yet. Therefore, we can still use the estimates for our planning optimization even when we have small number of queries.

To validate \textbf{H2}, we use the road network from \cite{biyik2018altruistic} and the equilibria benchmarks we developed: NE, BNE and BFNE with full flexibility. Here, we give the average latencies for the $4$-road network from that study which we visualize in Fig.~\ref{fig:wafr_network}, and where $\demandhum=0.4, \demandaut=1.2$ cars per second: (NE: $400.00$ sec, BNE: $125.66$ sec, BFNE (full flexibility): $102.85$ sec).

We then assumed we perfectly learned the preferences of the $5$ simulated users. We ran the planning optimization with $\bar{P}=0$ and $3$ different $\theta$ to show the trade-off. The results are summarized in Table~\ref{tab:wafr_new}.

\begin{table}[H]
	\caption{Results of Routing Simulation}
	\vspace{-5px}
	\label{tab:wafr_new}
	\centering
	\begin{tabular}{ccc}
		\hline
		$\theta$ & Avg. Latency (seconds) & Flow (cars/second) \\\hline
		$1$ & $90.41$ & $0.4412$ \\
		$20$ & $97.03$ & $1.2746$ \\
		$10^6$ & $111.28$ & $1.5964$ \\\hline
	\end{tabular}
	\vspace{-10px}
\end{table}

It can be seen we can adjust the trade-off between average latency and the served flow by tuning $\theta$. Also, given the human preferences, even when we served (almost) all of autonomous demand, our framework outperforms BNE. This shows its effectiveness on creating flexibility and supports \textbf{H2}.

For \textbf{H3}, we recruited $21$ subjects ($9$ female, $12$ male) with an age range from 19 to 60. In the first phase of the experiment, each participant was asked $40$ actively synthesized queries ($4$ roads + $1$ walking option). We then used their responses to get the maximum likelihood sample $(\boldsymbol{\bar\omega},\bar{\zeta})$. Afterwards, we designed $5$ different road networks each with $4$ different roads and an additional route where people may choose to walk. The $5$ different networks cover a range of different road lengths from $1.8$ kilometers to $78$ kilometers. For each network, we also set different $\theta$, $\demandhum$, $\demandaut$, and $\bar{P}$. By assuming all autonomous flow is in the service of these $21$ subjects, or of the groups that match with their preferences, we locally solved the planning optimization to get the pricing scheme for each traffic network. We refer to these results as \emph{anticipated} values.

In the second phase of the user study, we presented the route-price pairs and the walking option to the same $21$ subjects. For each of the $5$ networks, they picked the option they would prefer. Using these responses, we allocated the autonomous flow into the roads. However, it is technically possible that more users select a road than its maximum flow. To handle such cases, we assumed extra flow is transferred to the roads with smaller latencies without making the users pay more. If that is not feasible, extra flow is transferred to the slower roads, without any discount. While these break the fairness constraint, it rarely happens and affects only a very small portion of the flow. After autonomous flows are allocated, human driven cars selfishly chose their routes in a way to minimize the overall average latency. We refer to the results of this allocation as \emph{actual} values.

Table~\ref{tab:user_study} compares the anticipated and the actual values. We report latencies in seconds, flows in cars per second, and profit is a rate value with the unit USD per second. In order to show how our framework incentivizes flexible behavior, we also added other benchmarks: two where the same flow as actual flow is routed (BNE1 and BFNE1) and two where all flow demand is routed, i.e. no walking option (BNE2 and BFNE2). While BNE and BFNE assume completely inelastic demand, using them as benchmarks under both the actual flow and the all flow demand gives us insights about the success of our framework, because the allocation of vehicles under the actual flow is comparable to the elastic demand case.

\begin{figure}[h]
	\centering
	\vspace{-10px}
	\includegraphics[width=0.95\columnwidth]{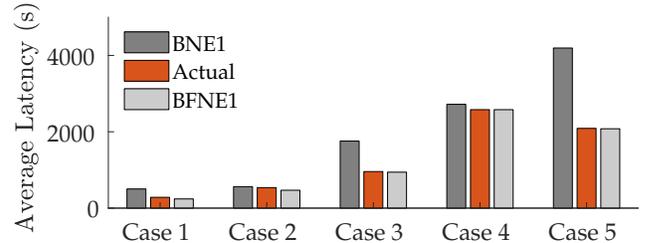}
	\caption{The comparison of the actual results and BNE1, both of which allocate the same amount of flow.}
	\label{fig:actual_vs_bne1}
	\vspace{-10px}
\end{figure}

It can be seen that there is generally an alignment between the anticipated and actual values. While the mismatch may be further reduced by doing more queries or having more users, the difference with BNE1 is significant. In all cases, our framework achieved to incentivize flexibility, which yielded lower average latencies compared to BNE1. Especially in Case 3 and Case 5, our framework approximately halved the average latency compared to the best Nash equilibria. In fact, our framework achieved a latency that is close to the lower bound set by BFNE1. We visualize these in Fig.~\ref{fig:actual_vs_bne1}. Furthermore, our framework successfully reduced flow demand when satisfying the full demand under selfishness is impossible (Case 1).

One caveat is the small amount of actual flow in Case 1, which also caused an important profit loss. This is because the roads are relatively shorter, and most users preferred walking over paying for an autonomous car. Our framework could not predict this, because the learned reward  functions failed to accurately estimate the probabilities. 

\begin{table*}[h]
	\caption{Results of Real-User Experiments}
	\label{tab:user_study}
	\centering
	\begin{tabular}{l|ccc|ccccc|ccc|}
		\cline{2-12}
		& \multicolumn{3}{c|}{\textbf{Anticipated}} & \multicolumn{1}{c}{\textbf{Actual}} &
		\multicolumn{2}{c}{\textbf{Pricing Scheme}} & \multicolumn{1}{c}{\textbf{BNE1}} &
		\multicolumn{1}{c|}{\textbf{BFNE1}} &
		\multicolumn{1}{c}{\textbf{Full}} & \multicolumn{1}{c}{\textbf{BNE2}} & \multicolumn{1}{c|}{\textbf{BFNE2}}  \\ \cline{2-12}
		& Flow & Av. Latency & Profit & Flow & Av. Latency & Profit & Av. Latency & Av. Latency & Flow & Av. Latency & Av. Latency \\ \hline 
		\multicolumn{1}{|l|}{Case 1} & $1.6622$ & $338.10$ & $11.11$ & $1.0429$ & $283.30$ & $3.85$ & $501.97$ & $240.80$ & $1.9000$ & Infeasible & $421.45$ \\
		\multicolumn{1}{|l|}{Case 2} & $1.5592$ & $510.97$ & $4.50$ & $1.5048$ & $537.30$ & $3.95$ & $562.20$ & $468.79$ & $1.6000$ & $562.20$ & $474.35$ \\
		\multicolumn{1}{|l|}{Case 3} & $1.2677$ & $921.06$ & $4.59$ & $1.3000$ & $957.71$ & $3.99$ & $1756.89$ & $941.81$ & $1.3000$ & $1756.89$ & $941.81$ \\
		\multicolumn{1}{|l|}{Case 4} & $1.5538$ & $2576.36$ & $40.01$ & $1.6000$ & $2579.52$ & $49.33$ & $2720.00$ & $2579.33$ & $1.6000$ & $2720.00$ & $2579.33$ \\
		\multicolumn{1}{|l|}{Case 5} & $1.2383$ & $2083.42$ & $48.48$ & $1.2048$ & $2089.53$ & $46.32$ & $4194.26$ & $2079.15$ & $1.3000$ & $4194.26$ & $2183.56$ \\ \hline
	\end{tabular}
	\vspace{2px}
	\caption*{For a fixed flow, BFNE serves as a lower bound on the latency. Achieving lower latency than BNE means we successfully incentivize flexible behavior.}
	\vspace{-25px}
\end{table*}

\section{CONCLUSION}\label{sct:conclusion}
In this work we address the efficiency of traffic networks with mixed autonomy. We develop a method of pricing rides with autonomous vehicles such that when a population chooses from these route and price options, and the human drivers choose the quickest routes available to them, the objective of decreasing travel latency and increasing road usage is achieved. To do so, we model how people choose between different route options with varying prices and latencies. Moreover, we develop a method for actively learning the parameters that describe the preferences of a population of users. We develop theoretical results which we use to gauge the performance of our algorithm, and conduct a user study showing that our method of parameterizing and actively learning the preferences of a human population is effective.

A wide horizon for further research remains. One could relax the assumption that the reward functions are linear to improve the prediction accuracy \cite{biyik2020active} and optimize for information gain in the active learning scheme, which can yield better data efficiency \cite{biyik2019asking}. Another direction is to consider an elastic demand for human drivers, as well as a variety of nondriving options including walking, biking, or taking the bus. In that case, $\zeta$ will have a multimodal distribution; we then need to learn the mixture.

More broadly, one can look at more general network topologies -- if route pricing is computed only with local information on each edge, pricing can make congestion worse when users have different price sensitivities \cite{brown2017fundamental}. One can also expand the model to include the role of information in decision making, as well as biases such as risk aversion. Through these future directions we can ensure the efficient operation of transportation networks that include autonomous vehicles. 

\bibliographystyle{IEEEtran}
\bibliography{refs}

\section{APPENDIX}\label{sct:appendix}
\noindent\textbf{Proof of Lemma~\ref{lma:free_flow}.} 
We begin by noting that for any given network and feasible flow demand, a \emph{continuum} of equilibria exist which satisfy the flow demand, where different equilibria have positive flow on different sets of roads. This fact stems from the two regimes (free-flow and congested) that exist, as well as the two different vehicle types. We note that for a congested road, a lower density yields a higher vehicle flow and lower latency. Accordingly, via the relationship with density, the latency function decreases smoothly with an increase in the flow of regular or autonomous cars. Note this does not mean that transferring flow to a road in a dynamic setting decreases its latency, rather that the function dictating the relationship between latency and flow in the congested regime is smoothly decreasing (in accordance with each quantity's relationship with the corresponding vehicle density).

With this in mind, we consider the continuum of equilibria that exist. Consider a Nash Equilibrium with positive flow on roads $\midcs$. Assume road $m$ is congested, as otherwise the lemma would be satisfied with $m'=m$. Roads ${[}m-1{]}$ must be congested so as to have the same latency as road $m$, as it is an equilibrium. We constructively find a different equilibrium as follows. Consider another configuration serving the same flow demand, where the flow on road $m$ serves less flow and roads  ${[}m-1{]}$ serve more flow as follows. Note that in each configuration serving the flow demand, vehicle flow is conserved but vehicle density is not. Accordingly, road $m$ has higher latency (from higher density) and roads ${[}m-1{]}$ have lower latency (from lower density). Since the flow-latency relationship in \eqref{eq:latency} has latency as a monotonically decreasing function with flow of each vehicle type, we can consider a new configuration where each road in ${[}m-1{]}$ has equal latency (which is less than the first equilibrium configuration) while serving more flow. Accordingly, we can construct one of the two following configurations.
\begin{enumerate}[nosep]
		\item the latencies on roads ${[}m-1{]}$ are reduced to $a_m$, the free-flow latency on road $m$, or
		\item road $m$ serves no flow and the latencies on roads ${[}m-1{]}$ are greater than $a_m$.
\end{enumerate}
In the first case, the lemma is satisfied with $m'=m$. In the second case, we can consider the same logic again, again considering a new configuration with less flow on road $m-1$ and more flow on roads  ${[}m-2{]}$. This continues until either we achieve case 1 above or until we are reduced to a single road. If that occurs, traffic can be routed in free-flow on that road, since any feasible flow in the congested regime is less than the maximum free-flow on a road. \qed

\noindent\textbf{Proof of Theorem~\ref{thm:BNE}.} The definition of Nash Equilibrium and the fact that latency on a road is always equal to or greater than its free-flow latency together imply that at Nash Equilibrium, if road $m$ has positive flow then all roads with free flow latency less than $a_m$ have positive flow as well. These also imply that if a road $m$ in free-flow has positive flow, roads with greater free-flow latency will have zero flow. Further, we use Lemma \ref{lma:free_flow} to show that all routings in the set of BNE will have one road in free-flow. Assume for the purposes of contradiction that we have a routing in the set of BNE in which only roads $\midcs$ have positive flow and all are congested, with equilibrium latency $\eqDelay$. The total cost is then $\eqDelay(\demandhum + \demandaut)$, where $\eqDelay > a_m$. By Lemma \ref{lma:free_flow}, another routing exists in the set of NE which uses roads $\mprimeidcs$, where $m' \le m$ and road $m'$ is in free flow. The cost of this equilibrium is $a_{m'}(\demandhum + \demandaut) \le \fflatency_m(\demandhum + \demandaut) < \eqDelay(\demandhum + \demandaut)$ contradicting our premise.
	
So far we have proved the numbered claims. We prove the remaining claim by contradiction. Assume there are two routings in $f$, $f' \in \text{BNE}(\demandhum, \demandaut)$ which have different free-flow roads, $m$ and $m'$ respectively. Assumption \ref{asmpt:ffl} implies $\fflatency_m \neq \fflatency_{m'}$; let $\fflatency_m < \fflatency_{m'}$. Since all selfish users experience the same latency, the total  latency of routing $f$ is $(\demandhum+\demandaut)\fflatency_m < (\demandhum+\demandaut)\fflatency_{m'}$, which is the total latency of routing $f'$. However by the definition of BNE, the total latency of the two routings are equal, yielding a contradiction. \qed

\noindent\textbf{Proof of Theorem~\ref{thm:BANE}.} The first property directly follows from Theorem \ref{thm:BNE}, as the regular vehicles have to be at a Nash equilibrium due to selfishness.
To prove the second property, we note that for roads that have higher latencies than road $\mstareq$, a Nash equilibrium is not necessary due to flexibility. As $\ell_i(\flowhum_i,\flowaut_i,s_i)$ is a non-increasing continuous function of $\flowaut_i$ and decreasing for $s_i=1$, roads that have higher latencies than road $\mstareq$ will always be in free-flow.

Now we assume some of the roads with indices greater than $\mstareq$ and less than $\mstarall$ are in free-flow, but not at maximum flow. Then we could simply transfer some flow from the road $\mstarall$ to those roads and have lower overall costs. This is a contradiction, completing the proof for the third property. \qed

\noindent\textbf{Proof of Theorem~\ref{thm:compute_BNE}.} 
Since road $\meq$, the longest equilibrium road, is in free-flow by Theorem \ref{thm:BNE} and $\mstareq$ is the minimum feasible $\meq$, our solution is restricted to the set of BNE. Accordingly, we can restrict our optimization to routing in which the longest equilibrium road is in free-flow. This allows us to write an optimization equivalent to checking the feasibility of a routing with the desired congestion profile:
\begin{align*}
&\max_{\flowhumvec, \flowautvec \in \mathbb{R}^N_{\ge 0}} 1 \\
&\quad \text{s.t.} \sum_{\pathidx \in \mstareqidcs}\flowhum_\pathidx = \demandhum, \; \, \sum_{i \in \mstareqidcs}\flowaut_\pathidx = \demandaut \\
&\qquad \; \, \latency_\pathidx(\flowhum_\pathidx,\flowaut_\pathidx,1) = \fflatency_\mstareq \; \forall \pathidx \in {[}\mstareq-1{]} \\
&\qquad \; \, \flowhum_\mstareq + \flowaut_\mstareq \le \capacity_\mstareq(\flowhum_\mstareq , \flowaut_\mstareq)
\end{align*}

The constraints can be shown to be affine in the decision variables. As it is a linear program, it can be solved in $O(N^3)$ time \cite{gonzaga1992path}. Finding $\mstareq$ requires a search in $O(N)$ time. \qed

\noindent\textbf{Proof of Theorem~\ref{thm:human_choice}.}
We first prove, similar to Luce's choice axiom \cite{luce2012individual}, changing the latency or the price of some roads does not alter the autonomous flow ratio between the other options, including the alternative option. For this, we look at $\mathbb{E}_{\boldsymbol{\omega},\zeta}\left[\frac{P(\pathidx_1 \mid \boldsymbol{\omega},\zeta,\latencyvec,\pricevec)}{P(\pathidx_2 \mid \boldsymbol{\omega},\zeta,\latencyvec,\pricevec)}\right]$, where $\pathidx_2$ is an undominated option, $P(\pathidx \mid \boldsymbol{\omega},\zeta,\latencyvec,\pricevec)$ is the probability of choosing option $\pathidx\in[N]\cup\{0\}$ under the given reward parameters, latency and price. We note
\begin{align*}
&P(\pathidx \mid \boldsymbol{\omega},\zeta,\latencyvec,\pricevec) = \frac{\exp(r(\latencyvec,\pricevec,\pathidx ; \boldsymbol{\omega}, \zeta))}{\sum_{\pathidx'=0}^N \exp(r(\latencyvec,\pricevec,\pathidx' ; \boldsymbol{\omega}, \zeta))} \; , \quad \text{so} \\
&\mathbb{E}_{\boldsymbol{\omega},\zeta}\left[\frac{P(\pathidx_1 \mid \boldsymbol{\omega},\zeta,\latencyvec,\pricevec)}{P(\pathidx_2 \mid \boldsymbol{\omega},\zeta,\latencyvec,\pricevec)}\right] = \mathbb{E}_{\boldsymbol{\omega},\zeta}\left[\frac{\exp(r(\latencyvec,\pricevec,\pathidx_1 ; \boldsymbol{\omega}, \zeta))}{\exp(r(\latencyvec,\pricevec,\pathidx_2 ; \boldsymbol{\omega}, \zeta))}\right] \; .
\end{align*}
As the reward of an option depends only on that option's price and latency, this proves the first statement above.


Equipped with this result, we now prove Theorem~\ref{thm:human_choice}. Assume the optimal solution is such that all human-driven flow is in congested roads. Let $k$ be the index of the highest free-flow latency road with nonzero human-driven flow. We first show there exists an equally optimal solution with no dominated roads in $[k]$. For this, we simply set the prices of roads $[k]$ such that they are all equal and the total autonomous demand in $[k]$ in the original and the new solution is the same. Since the new solution has no dominated roads in $[k]$, their new price has to be at least as high as the undominated roads in $[k]$ of the original solution, which implies the profit constraint is still satisfied. As the total demand served and the overall latency values are the same between these two solutions by the first statement above, the two solutions are equally good. In the remaining of the proof, we refer to the new solution as the ``optimal solution" for clarity, and show there exists a better solution, leading to a contradiction.
	
Denote the optimal solution with $(\latencyvec^*, \pricevec^*, {\flowhumvec}^*, {\flowautvec}^*)$, and its dominated options with $D^*$, noting $[k]\cap D^*\!=\!\emptyset$. Let the ratio of autonomous service users in road $i$ to the autonomous service users who decline the service be $\beta_i$, i.e., $\beta_i = \frac{{\flowaut}^*_{\pathidx}}{\demandaut - \sum_{\pathidx'\in[N]} {\flowaut}^*_{\pathidx'}}$.

Let $k'\!=\!\argmax_\pathidx \fflatency_\pathidx$ subject to $\pathidx\!\leq\! k$ and $\left(\sum_{\pathidx'\in [k]} {\flowhum}^*_{\pathidx'}, \sum_{\pathidx'\in [k]}{\flowaut}^*_{\pathidx'}\right)$ can be allocated into $[k']$ when all roads in $[k']$ have latency equal to $\fflatency_{k'}$. Existence of such a $k'$ is guaranteed by Lemma~\ref{lma:free_flow}.

Using $k'$, we propose an alternative solution $(\latencyvec',\!\pricevec',\!{\flowhumvec}',\!{\flowautvec}')$:
\begin{align*}
\latency'_{\pathidx} &= \begin{cases}
\latency^*_{\pathidx} & \text{ if } \pathidx\in[N]\setminus[k]\\
a_i & \text{ if } \pathidx\in[k]\setminus[k']\\
\fflatency_{k'} & \text{ if } \pathidx\in[k']
\end{cases} \; , \\
\price'_{\pathidx} &= \begin{cases}
\price^*_{\pathidx} & \text{ if } \pathidx\in[N]\setminus\left([k] \cup D^*\right)\\
\price^*_{\pathidx} + \epsilon & \text{ if } \pathidx\in[k] \cup D^*
\end{cases} \; .
\end{align*}
where $\epsilon \geq 0$ is such that the ratio of autonomous service users in $[k']$ to the autonomous service users who decline the service in the alternative solution is equal to $\sum_{\pathidx\in[k]} \beta_\pathidx$. Since $\latency'_\pathidx < \latency^*_\pathidx$ for $\forall \pathidx\in[k']$, $\omega_2>0$ for all users, and $[k]\cap D^* = \emptyset$, the existence of such an $\epsilon$ is guaranteed.

By the first statement, we know the ratios ($\beta_\pathidx$'s) in the alternative solution will be equal to the optimal solution for roads in $[N]\setminus ([k] \cup D^*)$ as their latencies and prices are the same. Similarly, roads in $D^*$ are also dominated in the alternative solution, because no road has higher latency and prices went up maximally (by $\epsilon$) for those roads. This means the ratios are equal between two solutions for all roads in $[N]\setminus [k]$. Noting roads in $[k]\setminus[k']$ are dominated in the alternative solution, as they have the same price as the roads in $[k']$ but higher latency, we write the total autonomous demand:
\begin{align*}
\demandaut = \left(\demandaut - \sum_{\pathidx\in[N]} {\flowaut}'_{\pathidx}\right)(1 + \beta_1 + \beta_2 + \dots + \beta_N)
\end{align*}
by the construction of $\epsilon$. As the total autonomous demand is the same between two solutions, we get $\demandaut - \sum_{\pathidx\in[N]} {\flowaut}'_{\pathidx} = \demandaut - \sum_{\pathidx\in[N]} {\flowaut}^*_{\pathidx}$ meaning the flow of users who decline the service is the same. This implies ${\flowaut}'_\pathidx = {\flowaut}^*_\pathidx$ for $\forall \pathidx \in [N]/[k]$, and $\sum_{\pathidx\in[k]} {\flowaut}'_\pathidx = \sum_{\pathidx\in[k]} {\flowaut}^*_\pathidx$. As the selection of $k'$ ensures there is enough room for selfish vehicles in $[k']$ in the alternative solution, we have $\sum_{\pathidx\in[k']} {\flowhum}'_\pathidx = \sum_{\pathidx\in[k]} {\flowhum}^*_\pathidx$.

Finally, the alternative solution satisfies the profit constraint because the price and the autonomous flow in $[N]\setminus [k]$ are the same as the optimal solution, and the remaining autonomous flow pays higher price in the alternative solution.

Overall, the alternative solution is feasible, serves the same amount of flow, but has lower latency overall\textcolor{blue}{,} $(\latencyvec^*, \pricevec^*, {\flowhumvec}^*, {\flowautvec}^*)$ cannot be the optimal solution. \qed

\noindent\textbf{Role of Assumption \ref{asmpt:ffl}.}
Without this assumption, $\mstareq$ in Theorem~\ref{thm:BNE} would instead represent a set of roads, all of which would be in free-flow. Theorem~\ref{thm:BANE} would change similarly to Theorem~\ref{thm:BNE}. Theorem~\ref{thm:compute_BNE} is not altered, the constraint associated with $\mstareq$ would instead apply to all roads within the set. Finally, Theorem~\ref{thm:compute_BANE} would remain the same as well. What would potentially change this theorem is if the set of roads in $\mstareq$ could have different congestion levels in the computed BFNE. To see why this is not the case, note that the flexibility profile is with respect to the quickest route available to users (\emph{i.e.} the least congested road in the set $\mstareq$). Accordingly, having another road in the set be more congested would not help satisfy any constraint related to the flexibility profile and would also not serve more flow than if it had the same latency as the minimum latency road in $\mstareq$. Hence, all roads in $\mstareq$ will have the same latency, so the computational complexity remains the same. Theorem~\ref{thm:human_choice} would still hold: there exists a free-flow road used by human drivers in the optimal solution.



\vspace{-85px}
\begin{IEEEbiography}[{\includegraphics[width=1in,height=1.25in,clip,keepaspectratio]{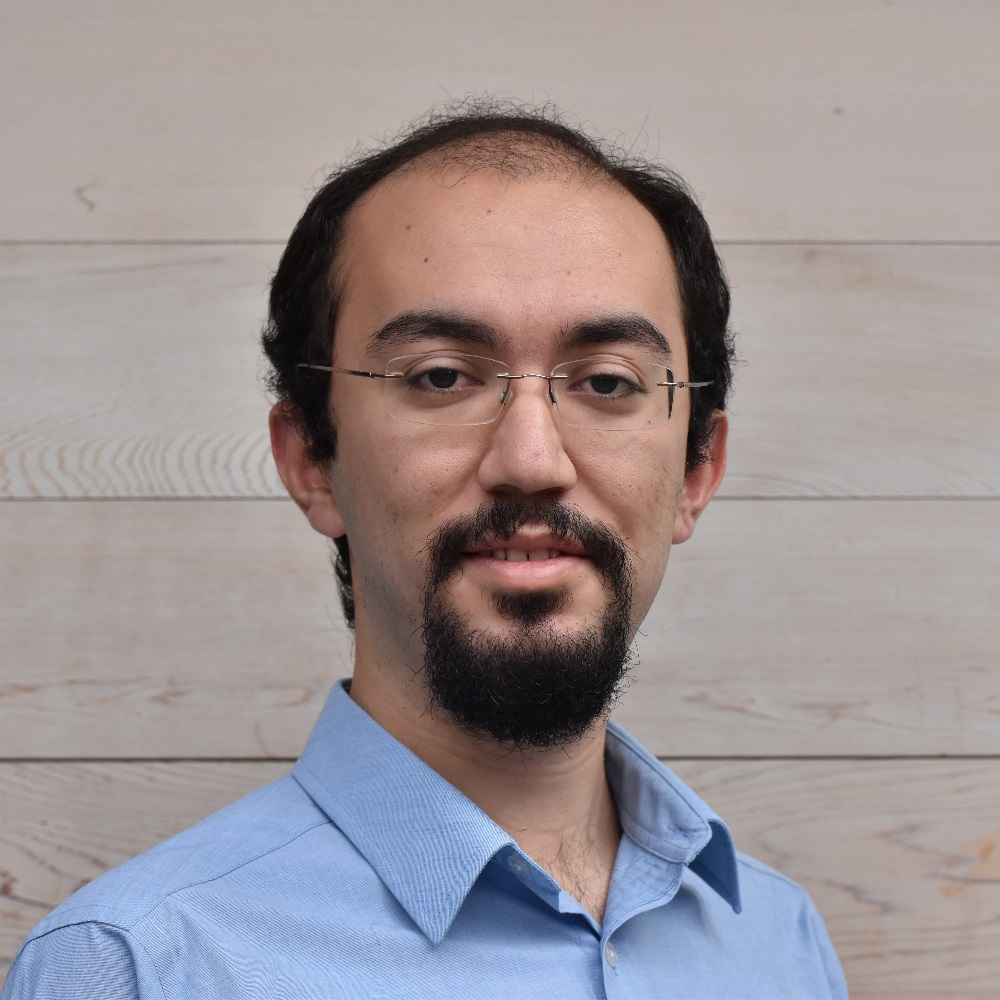}}]{\\Erdem B\i y\i k}
	is a fourth-year Ph.D. candidate in the EE Department at Stanford. He received his B.Sc. degree in electrical and electronics engineering from Bilkent University, Turkey, in 2017. He is interested in enabling robots to learn from various forms of human feedback, and designing robot policies to improve the performance of multi-agent systems both in cooperative and competitive settings.
	\vspace{-90px}
\end{IEEEbiography}

\begin{IEEEbiography}[{\includegraphics[width=1in,height=1.25in,clip,keepaspectratio]{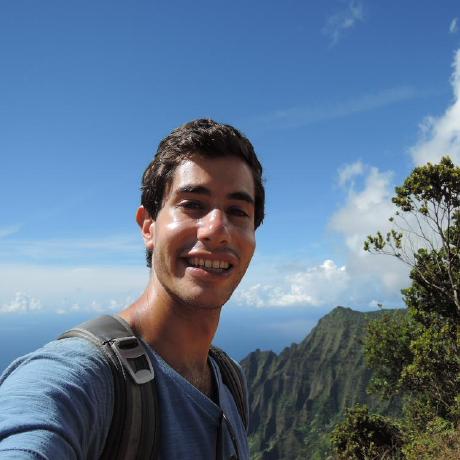}}]{\\Daniel A. Lazar}
	is a fifth-year Ph.D. candidate in the ECE Department at the University of California, Santa Barbara. In 2014, he received a B.Sc. in Electrical Engineering from Washington University in St. Louis, after which he spent two years working in wireless communications. His research focuses on control of transportation networks.
	\vspace{-85px}
\end{IEEEbiography}

\begin{IEEEbiography}[{\includegraphics[width=1in,height=1.25in,clip,keepaspectratio]{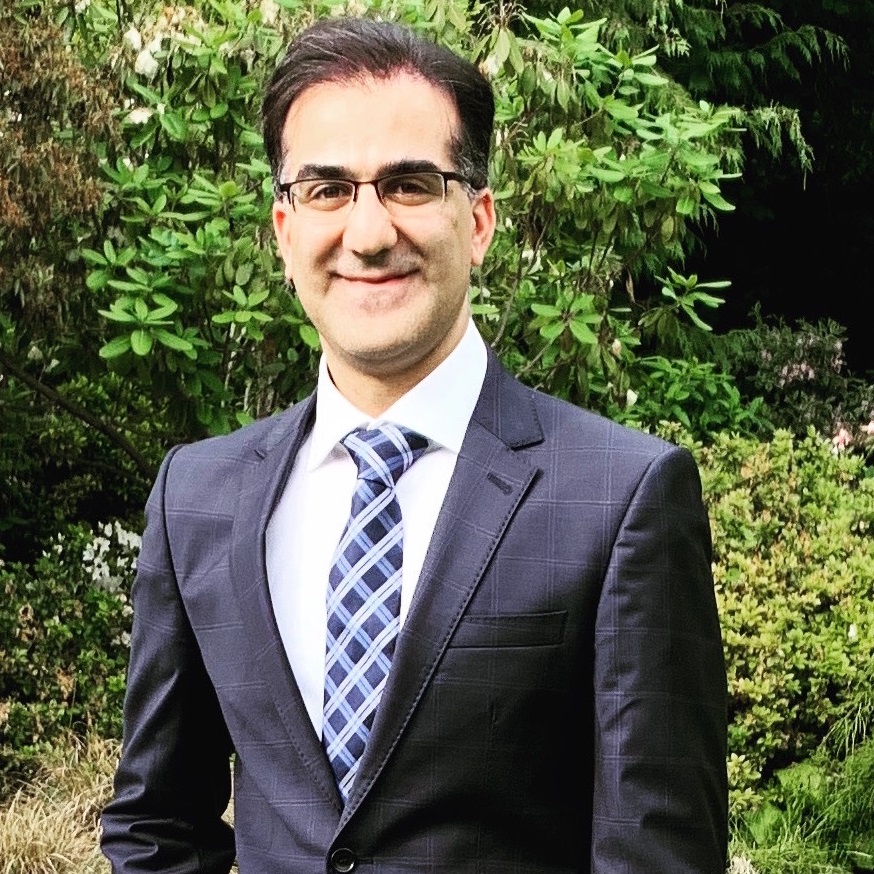}}]{Ramtin Pedarsani}
	is an Assistant Professor in the ECE Department at the University of California, Santa Barbara. He received the B.Sc. degree in electrical engineering from the University of Tehran in 2009, the M.Sc. degree in communication systems from the Swiss Federal Institute of Technology (EPFL) in 2011, and his Ph.D. from the University of California, Berkeley, in 2015. His research interests include networks, game theory, machine learning, and transportation systems.
	\vspace{-90px}
\end{IEEEbiography}

\begin{IEEEbiography}[{\includegraphics[width=1in,height=1.25in,clip,keepaspectratio]{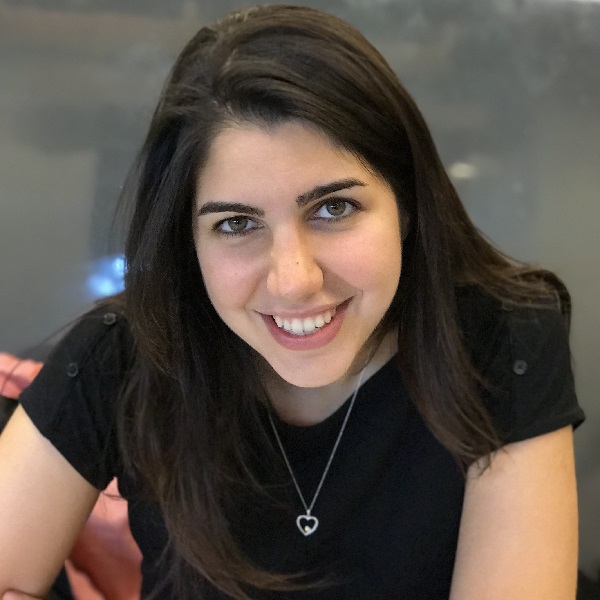}}]{\\Dorsa Sadigh}
	is an Assistant Professor in the CS and EE departments at Stanford University. Her research interests lie in the intersection of robotics, learning and control theory. Specifically, she is interested in developing algorithms for safe and adaptive human-robot interaction. Dorsa has received her doctoral degree in Electrical Engineering and Computer Sciences (EECS) at UC Berkeley in 2017, and has received her B.Sc. in EECS at UC Berkeley in 2012.
	\vspace{-50px}
\end{IEEEbiography}


\end{document}